\documentclass[fleqn,usenatbib]{mnras}

\usepackage{mathptmx}
\usepackage{threeparttable}
\usepackage{color}
\usepackage{adjustbox}
\usepackage{relsize}
\definecolor{sig_col}{rgb}{0.0, 0.0, 1.0}
\newcommand{\sig}[1]{\textcolor{sig_col}{#1}}
\definecolor{sig2}{rgb}{0.0, 0.44, 1.0}

\usepackage{hyperref}

\usepackage[T1]{fontenc}
\usepackage{soul}

\DeclareRobustCommand{\VAN}[3]{#2}
\let\VANthebibliography\thebibliography
\def\thebibliography{\DeclareRobustCommand{\VAN}[3]{##3}\VANthebibliography}

\usepackage{graphicx}	
\usepackage{amsmath}	
\usepackage{amssymb}	
\usepackage{mathtools}
\usepackage{float}
\usepackage{graphicx}
\usepackage{ulem}
\usepackage{lscape}
\usepackage{xspace}

\newcommand{\GG}[1]{}

\title[Asynchronous accretion can mimic diverse white dwarf pollutants II]{Asynchronous accretion can mimic diverse white dwarf pollutants II: water content}

\author[Brouwers Et Al.]{
	Marc G. Brouwers$^{1}$\thanks{E-mail: mgb52@cam.ac.uk},
	Andrew M. Buchan$^{1}$, Amy Bonsor$^{1}$, Uri Malamud$^{2,3}$, Elliot Lynch$^{4}$\newauthor Laura Rogers$^{1}$, Detlev Koester$^{5}$ \\
	$^{1}$Institute of Astronomy, University of Cambridge, Madingley Road, Cambridge CB3 0HA \\
	$^{2}$Department of Physics, Technion - Israel Institute of Technology, Technion City, 3200003 Haifa, Israel\\
	$^{3}$School of the Environment and Earth Sciences, Tel Aviv University, Ramat Aviv, 6997801 Tel Aviv, Israel\\
	$^{4}$ Univ Lyon, Univ Lyon1, Ens de Lyon, CNRS, Centre de Recherche Astrophysique de Lyon UMR5574, F-69230, Saint-Genis,-Laval, France.\\
	$^{5}$ Institut f{\"u}r Theoretische Physik und Astrophysik, University of Kiel, D-24098 Kiel, Germany
}

\date{Accepted 8 Nov 2022. Received 28 Oct 2022; in original form 18 Aug 2022}

\pubyear{2022}

\begin{document}
	\maketitle
	
	\begin{abstract}
	    Volatiles, notably water, are key to the habitability of rocky planets. The presence of water in planetary material can be inferred from the atmospheric oxygen abundances of polluted white dwarfs, but this interpretation is often complex. We study the accretion process, and find that ices may sublimate and accrete before more refractory minerals reach the star. As a result, a white dwarf's relative photospheric abundances may vary with time during a single accretion event, and do not necessarily reflect the bulk composition of a pollutant. We offer two testable predictions for this hypothesis: 1. cooler stars will more often be inferred to have accreted wet pollutants, and 2. there will be rare occurrences of accretion events with inferred volatile levels far exceeding those of pristine comets. To observationally test these predictions, we statistically constrain the water content of white dwarf pollutants. We find that in the current sample, only three stars show statistically significant evidence of water at the $2\sigma$ level, due to large typical uncertainties in atmospheric abundances and accretion states. In the future, an expanded sample of polluted white dwarfs with hydrogen-dominated atmospheres will allow for the corroboration of our theoretical predictions. Our work also shows the importance of interpreting pollutant compositions statistically, and emphasizes the requirement to reduce uncertainties on measured abundances to allow for statistically significant constraints on their water content. 
	\end{abstract}
	
	\begin{keywords}
		white dwarfs – planetary systems – transients: tidal disruption events – planet–disc interactions
	\end{keywords}
	

\section{Introduction}\label{sect:introduction}
White dwarfs are the dense, degenerate remnants of stars that have lost their envelopes during post-main sequence evolution. High surface gravity leads to chemical stratification of their outer layers, with elements heavier than H/He sinking out of sight within days to millions of years, depending on stellar type and age \citep{Koester2009}. Nevertheless, up to 50\% of white dwarfs have atmospheres that are polluted with heavy elements \citep{Zuckerman2003, Zuckerman2010, Koester2014}, evidence that they experienced recent accretion of planetary material \citep{Jura2003, Jura2014, Veras2021b}. Ongoing accretion has independently been confirmed for one system via the detection of X-rays \citep{Cunningham2022}.

If multiple elements are detected in a white dwarf atmosphere, their relative abundances can be used to constrain the composition of the pollutants \citep[e.g.,][]{Hollands2018, Harrison2018, Harrison2021b, Swan2019, Putirka2021, Buchan2022}. Notably, the detection of oxygen together with the other major rock-forming elements (Fe, Si, and Mg) makes it possible to infer the water content of the accreted material \citep{Klein2010, Farihi2013}. The delivery of comets containing water may be crucial for habitable planet formation in dry inner regions \citep{Morbidelli2000, Albarede2009, Raymond2009}, and polluted white dwarfs offer a direct way to investigate the presence of water-rich material around other stars. 

\begin{figure*}
\centering
\includegraphics[width=\hsize]{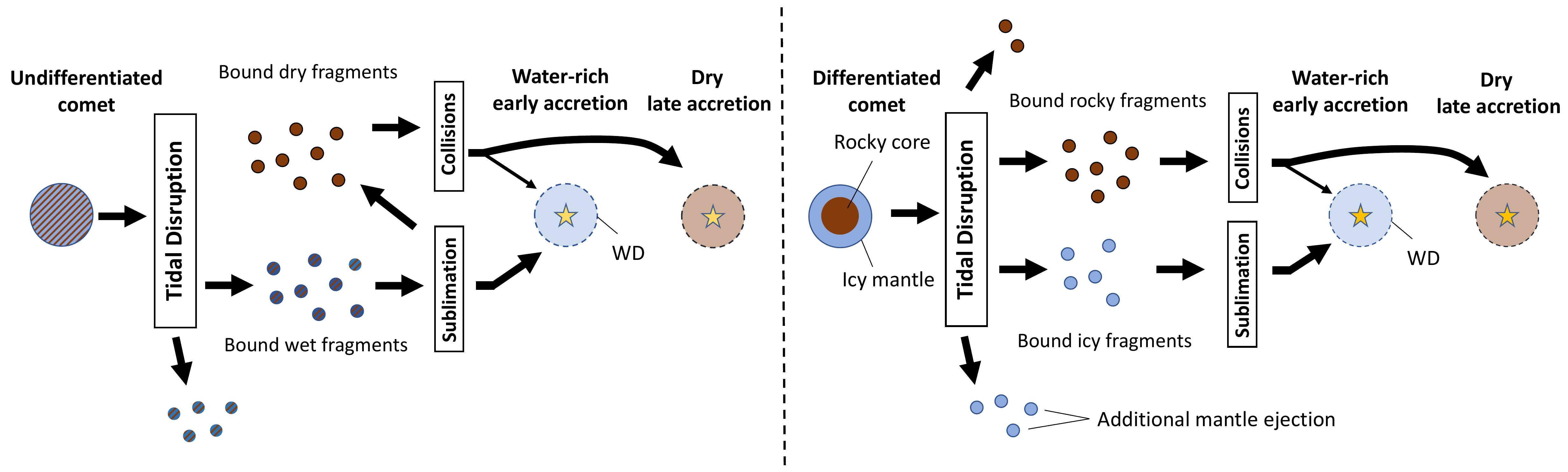} 
\caption{Schematic overview of how the asynchronous accretion of an ice-rich comet onto a white dwarf can appear either volatile-rich or dry, depending on when the system is observed. Left panel: an undifferentiated comet tidally disrupts, producing wet fragments. If these fragments dry out before they collide, a short water-rich accretion phase is followed by a longer, dry accretion phase. Right panel: an ice-rock differentiated comet tidally disrupts, producing both icy and dry fragments. The icy fragments are ejected in greater proportions (see paper I). The bound ice rapidly sublimates - before most fragments catastrophically collide and accrete, again leading to a short, volatile-rich early accretion phase and a longer, dry, second accretion phase. An accompanying version of this figure that details asynchronous core-mantle accretion is shown in paper I. \label{fig:schematic}}
\end{figure*}

Analyses of individual white dwarfs indicate that, while most pollutants are made up of dry minerals \citep{Gaensicke2012}, nearly one in four systems with oxygen detections is inferred to contain a substantial amount of water ice \citep{Farihi2013, Farihi2016, Raddi2015, Xu2017, Xu2019, Hoskin2020, Klein2021, Hollands2022}. Independently, volatile accretion has also been investigated using white dwarfs with helium-dominated atmospheres, where trace amounts of hydrogen record the cumulative accretion of water ice throughout their history \citep{Veras2014b, GentileFusillo2017, Izquierdo2018, Izquierdo2021}. However, in the population studies by \citet{Jura2012, Jura2014} and \citet{GentileFusillo2017}, water is found to be far more rare than indicated by the analysis of individual systems, with a fraction below 1\% of the total metal mass accreted in the form of water. The origin of this apparent mismatch is not currently well understood, but could hint at the over-interpretation of water in the compositional analysis of individual white dwarfs.

Until now, studies that infer the composition of white dwarf pollutants make the assumption that the different parts of an object (e.g., core/mantle, volatile/refractory) enter the white dwarf in a constant proportion, such that ongoing accretion is always representative of its bulk content. In this work, we expand upon an accompanying paper (\citealt{Brouwers2022b}, henceforth paper I) to investigate the possibility that accretion occurs \textit{asynchronously}, with some parts of a pollutant accreting faster than others. In particular, we study the scenario suggested by \citet{Malamud2016}, who hypothesized that the ices contained in a pollutant can sublimate and accrete before more refractory minerals reach the star. In this scenario, where sublimative erosion outpaces alternative accretion processes such as scattering, collisional grind-down and drag forces \citep[e.g.][]{Veras2015b, Swan2021, Li2021, Malamud2021, Brouwers2022}, the ice and rock in a pollutant will largely accrete separately, and the photospheric abundances of white dwarfs do not necessarily reflect the bulk composition of their pollutants. We focus on defining observational predictions that result from this form of asynchronous accretion, and collate a white dwarf sample from the literature to test these predictions. In our compositional analysis, we confront the problem that to accurately infer the composition of accreting material, the star's accretion state (i.e. build-up/steady-state/declining) needs to be well-constrained. We demonstrate a procedure where a Bayesian model (\texttt{PyllutedWD}, \citealt{Harrison2021b, Buchan2022}) is used to quantify the uncertainties on the accretion state, with allows for a statistical constraint on the presence of water in the accreted material. Our results highlight the difficulty of making statistically strong statements regarding the composition of pollutants in white dwarfs, especially those with helium-dominated atmospheres, where accretion states are often subject to large uncertainties.

This paper is organized as follows. We first compute the sublimative erosion of icy and rocky fragments in Section \ref{sect:sublimation}, and show how asynchronous accretion can affect white dwarf abundances in Section \ref{sect:evolution_abundances}. To test the predictions of asynchronous accretion, we analyse the \textit{oxygen excess} of white dwarf pollutants in Section \ref{sect:observational_comparison}. Finally, we discuss the results in Section \ref{sect:discussion} and conclude in Section \ref{sect:conclusions}.

\section{Timescale of sublimative erosion}\label{sect:sublimation}
The accretion of planetary material onto a white dwarf is usually thought to begin with the tidal disruption of a pollutant \citep{Debes2012a, Veras2014, Nixon2020, Malamud2020a, Malamud2020b}, forming a highly eccentric disc with its fragments. If these fragments face intense radiation from the white dwarf, they can begin to lose their material to sublimation, similar to the fate of comets near the Sun \citep{Binzel2004}. The timescale for the sublimative erosion of a fragment can be estimated from the integrated stellar flux over a complete orbit, defined by Kepler's orbital equations of distance to a central star ($r$) with mass $M_\mathrm{WD}$:
\begin{equation} \label{eq:r_v}
    r = \frac{a(1-e^2)}{1-e \;\mathrm{cos}(\theta)} \;,\; \frac{d\theta}{dt} = \left[\frac{G M_\mathrm{WD}}{a^3\left(1-e^2\right)^3}\right]^{\frac{1}{2}}\left(1-e\; \mathrm{cos}(\theta)\right)^2,
\end{equation}
where $\theta$ is the true anomaly (with the pericentre at $\theta=\pi$), and $a,e$ are the fragment's semi-major axis and eccentricity. At any point on the orbit, the flux of stellar radiation through its surface ($J_\mathrm{WD}$) averages to:
\begin{equation}\label{eq:incident_flux}
    J_\mathrm{WD} = \frac{(1-A) \sigma_\mathrm{sb} R_\mathrm{WD}^2 T_\mathrm{WD}^4}{4 r^2},
\end{equation}
where $\sigma_\mathrm{sb}$ is the Stephan-Boltzmann constant, $A$ is the fragment's albedo, and $R_\mathrm{WD}, T_\mathrm{WD}$ are the stellar radius and temperature. Some of this flux is used to heat or sublimate the fragment, while a portion is re-radiated back to space. For ices with very low sublimation temperatures, the importance of thermal re-radiation is negligible, allowing for the simplification that all the incident flux is used for heating and sublimation. With this assumption, the shrinkage of a fragment's radius due to sublimative erosion ($dR_\mathrm{frag}/dt$) can be time-averaged over a complete orbit with period $P_\mathrm{frag} = 2\pi \sqrt{a_\mathrm{frag}^{3}/(G M_\mathrm{WD})}$, and is:
\begin{subequations}
\begin{align}
    \bar{\frac{dR_\mathrm{frag}}{dt}} &= -\frac{1}{P_\mathrm{frag}}
    \int_{0}^{2 \pi} \frac{ J_\mathrm{WD}}{H_\mathrm{sub}\rho_\mathrm{frag}} \left(\frac{d\theta}{dt}\right)^{-1} d\theta \\
    &= -\frac{(1-A) \sigma_\mathrm{sb} T_\mathrm{WD}^4 R_\mathrm{WD}^2}{4 H_\mathrm{sub} \rho_\mathrm{frag} a_\mathrm{frag}^2 \sqrt{1-e_\mathrm{frag}^2}}. \label{eq:dRdt}
\end{align}
\end{subequations}
where $H_\mathrm{sub}$ is the enthalpy of sublimation. Note that Eq. \ref{eq:dRdt} is independent of the fragment's size. With this expression, the timescale for sublimative erosion $t_\mathrm{sub}$ of icy fragments follows as:
\begin{subequations}
\begin{align}
    t_\mathrm{sub} &= -\frac{R_\mathrm{frag}}{\bar{dR_\mathrm{frag}/dt}} \\
    &= \frac{4 H_\mathrm{sub} \rho_\mathrm{frag}a_\mathrm{frag}^2\sqrt{1-e_\mathrm{frag}^2} R_\mathrm{frag}}{(1-A)\sigma_\mathrm{sb} T_\mathrm{WD}^4 R_\mathrm{WD}^2}\\
    &\simeq 6.9 \cdot 10^4 \; \mathrm{yr} \left(\frac{T_\mathrm{WD}}{2\cdot10^4 \; \mathrm{K}}\right)^{-4} \left(\frac{R_\mathrm{WD}}{\mathrm{R_\oplus}}\right)^{-2} \left(\frac{H_\mathrm{sub}}{2.7\cdot 10^{10}\;\mathrm{erg/g}}\right)  \label{eq:t_sub} \\
    & \qquad \left(\frac{r_\mathrm{B}}{\mathrm{R_\odot}}\right)^\frac{1}{2} \left(\frac{a_\mathrm{frag}}{3\;\mathrm{AU}}\right)^\frac{3}{2}
    \left(\frac{\rho_\mathrm{frag}}{1 \; \mathrm{g/cm^3}}\right)
    \left(\frac{1-A}{1}\right)^{-1}
    \left(\frac{R_\mathrm{frag}}{\mathrm{km}}\right), \nonumber
\end{align}
\end{subequations}
\begin{table}
\caption{Mechanical and thermal properties of forsterite ($\mathrm{Mg_2SiO_4}$), iron ($\mathrm{Fe}$), and crystalline water ice ($\mathrm{H_2O}$).}
\label{table_parameters}
\centering
\addtolength{\tabcolsep}{-0.9 mm}
\begin{tabular}{l c c c}
\hline
\hline
Symbol & Forsterite ($\mathrm{Mg_2SiO_4}$) & Iron ($\mathrm{Fe}$) & Ice ($\mathrm{H_2O}$)\\ 
\hline
$\mu^{(1)}$ & 140.69 & 55.85 & 18.02 \\
$\rho^{(2)}$ & 3.27 & 7.87 & 1 \\
$C_1^{(3)}$ & 34.1$^{(a,b)}$ & 29.2$^{(a,c)}$ & 31.1$^{(d)}$ \\
$C_2^{(4)}$ & 65308$^{(a,b)}$ & 48354$^{(a,c)}$ & 6135$^{(d)}$ \\
$H_\mathrm{sub}^{(5)}$ & $8.1 \cdot 10^{10} {}^{(e)}$ & $8.3 \cdot 10^{10} {}^{(e)}$ & $2.7\cdot10^{10} {}^{(f,\star)}$ \\
$\eta_\mathrm{sub}^{(6)}$ & 0.2$^{(g,\dagger)}$ & 1$^{(a)}$ & 0.2$^{(d,h)}$ \\
\hline
\end{tabular}
\begin{tablenotes}
\item Parameters: $^{(1)}$ molecular weight [$\mathrm{m_u}$] $^{(2)}$ density [$\mathrm{cm^3\;g^{-1}}$] $^{(3)}$ vapour pressure constant [no dim] $^{(4)}$ vapour pressure constant [K] $^{(5)}$ enthalpy of sublimation (heating + vaporization) [$\mathrm{erg\;g^{-1}}$] $^{(6)}$ kinetic inhibition [no dim]
\item References: $^{(a)}$ \citet{VanLieshout2014} $^{(b)}$ \citet{Nagahara1994} $^{(c)}$ \citet{Ferguson2004} $^{(d)}$ \citet{Gundlach2011} $^{(e)}$ \citet{Podolak1988} $^{(f)}$ \citet{Huebner2006} $^{(g)}$ \citet{Steckloff2021} $^{(h)}$ \citet{Beckmann1982}\\
\item \textbf{Notes:} $^{(\star)}$ evaluated at an initial comet temperature of 0 K$^{(\dagger)}$ evaluated at 2400 K
\end{tablenotes}
\end{table}
\begin{figure}
\centering
\includegraphics[width=\hsize]{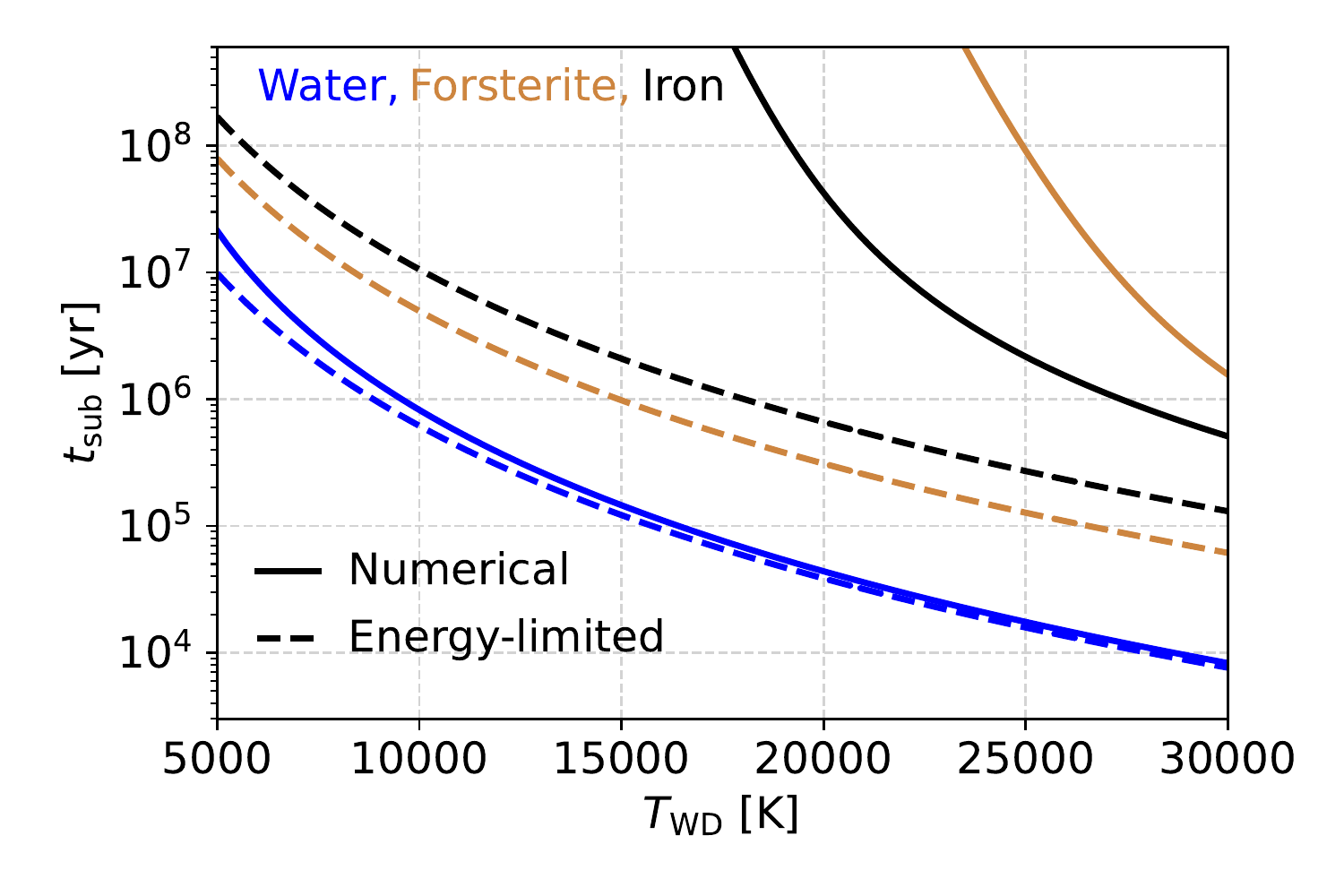}
\caption{Numerically calculated timescales for the sublimative erosion of large, km-sized fragments (solid lines), compared to the energy-limited approximation (dashed lines, Eq. \ref{eq:t_sub}). Fragments composed of water ice (blue) sublimate efficiently within $10^4-10^6$ yr around warm stars, and the two curves largely overlap. Fragments composed of refractory species like forsterite (brown) and iron (black) only begin to sublimate around the hottest white dwarfs, and re-radiate most of the incident stellar flux back to space, causing a divergence from the energy-limited curves.
\label{fig:sublimation_km}}
\end{figure}
where in the last line we used that $1-e_\mathrm{frag}^2 \simeq 2r_\mathrm{B}/a_\mathrm{frag}$ because $1-e_\mathrm{frag}=r_\mathrm{B}/a_\mathrm{frag} \sim O(\mathrm{R_\odot}/\mathrm{AU}) \ll 1$ in an eccentric tidal disc. The value $r_\mathrm{B}$ refers to the breakup distance of the asteroid, assumed equal to the pericentre of its orbit ($r_\mathrm{B}=a_\mathrm{frag}(1-e_\mathrm{frag})$). The largest fragments expected in the tidal disc are unlikely to exceed 1 km in size \citep{Zhang2021, Brouwers2022}, and might be significantly smaller if the inherent strength of the ice is sub-kP \citep{Greenberg1995, Davidsson1999, Gundlach2016}. Eq. \ref{eq:t_sub} indicates that even the largest icy fragments are expected to sublimate fully within 1 Myr if the stellar temperature exceeds $10^4$ K, and much faster if the star is warmer or the fragment is smaller. Around hot stars, the sublimation timescale is short compared to the typical duration of accretion events, as inferred from the statistical differences between hydrogen and helium white dwarfs \citep{Girven2012, Cunningham2021}. In this case, icy fragments are expected to completely sublimate before the accretion process of more refractory components has completed.

\subsection{Very slow sublimation of refractory minerals }\label{sect:numerical}
To contrast the sublimative erosion of ices and more refractory species, we formulate a simple numerical model that accounts for the fragment's thermal re-radiation into space. The fragments are assumed to have uniform compositions, consisting either of core-like material (iron), mantle-like material (forsterite), or ice (water). The sublimation parameters for these materials are shown in Table \ref{table_parameters}. Furthermore, it is assumed that the heating of the fragments proceeds from a thin, hot outer layer, with negligible internal heat transport (see Appendix \ref{appendix:conduction}).

The energy flux through the fragment's surface contains terms for the incident stellar radiation ($J_\mathrm{WD}$, Eq. \ref{eq:incident_flux}), as well as the re-radiation ($J_\mathrm{emit}$) and sublimation ($J_\mathrm{sub}$) from the fragment's surface:
\begin{subequations}
\begin{align}
    J_\mathrm{WD} &= J_\mathrm{emit} + J_\mathrm{sub} \\
    \frac{(1-A)\sigma_\mathrm{sb}  R_\mathrm{WD}^2 T_\mathrm{WD}^4}{4 r^2} &=  \sigma_\mathrm{sb} T_\mathrm{s}^4 + H_\mathrm{sub} I_\mathrm{sub}, \label{eq:sublimation_energy_conservation}
\end{align}
\end{subequations}

where $T_\mathrm{s}$ is the surface temperature, and both the fragment and the star are assumed to emit as perfect black bodies. The mass flux $I_\mathrm{sub}$ at the surface should account for the processes of sublimation and re-condensation. The kinetic theory of gases gives the following equation \citep{Knudsen1909, Langmuir1913}:
\begin{equation}\label{eq:massloss}
    I_\mathrm{sub} =  \eta_\mathrm{sub}P_\mathrm{vap}\left(\frac{\mu}{2\pi k_\mathrm{b} T_\mathrm{s}}\right)^\frac{1}{2},
\end{equation}
with saturated vapour pressure $P_\mathrm{vap}$. The kinetic inhibition $\eta_\mathrm{sub}$ describes the divergence of the sublimation from being the perfect inverse of condensation \citep{Kossacki1999}, and has to be determined experimentally. The same is true for the vapour pressure, which typically follows the empirical Clausius-Clapeyron relation \citep{Clausius1850}:
\begin{equation}
    P_\mathrm{vap} = \mathrm{exp}\left(C_1 - \frac{C_2}{T_\mathrm{s}}\right) \; \mathrm{dyne/cm^2},
\end{equation}
with constants $C_1, \;C_2$ (see Table \ref{table_parameters}). We solve for the temperature of the fragment's outer layer from Eq. \ref{eq:sublimation_energy_conservation} with a root-finding procedure at every time-step across a full orbit, and calculate the mass loss via Eq. \ref{eq:massloss}. The fragment's albedo is set equal to the canonical value of 0.04 for comets \citep{Bernardinelli2021}. To approximate the complete disintegration of a fragment, its sublimative mass-loss is integrated over one complete orbit, and this result is extrapolated, using the observation that sublimative shrinkage (of a fragment's radius) is independent of its remaining size.

The numerically calculated timescales for sublimative erosion are plotted in Fig. \ref{fig:sublimation_km}, with comparison to the energy-limited expression (Eq. \ref{eq:t_sub}). The values for water ice accurately follow the analytical expression, while the two curves diverge for fragments composed of forsterite or iron. These refractory fragments lose the vast majority of their mass near the pericentre of their eccentric orbits, where they only spend a tiny fraction of their time. At distances further from the star, nearly all the stellar flux is re-radiated without substantial sublimation. Even with stellar temperatures of 20,000 K, the sublimative erosion of km-sized iron and forsterite fragments in a tidal disc requires more than $10^7$ yr. Therefore, we find that while ices can rapidly sublimate around hot white dwarfs, fragments made of more refractory species are expected to remain intact until they are collisionally ground down to dust. 

\section{Implied water content during accretion}\label{sect:evolution_abundances}
In the preceding section, we showed that icy fragments can face rapid sublimative erosion in the tidal discs where they are released. If sublimative erosion outpaces other accretion channels, most of the volatile ices will accrete onto the white dwarf before more refractory species arrive. In this section, we show how such an asynchronicity causes the star's relative photospheric abundances to vary over time during a single accretion event, which in turn affects the implied composition of its pollutants.

\begin{figure}
\centering
\includegraphics[width=\hsize]{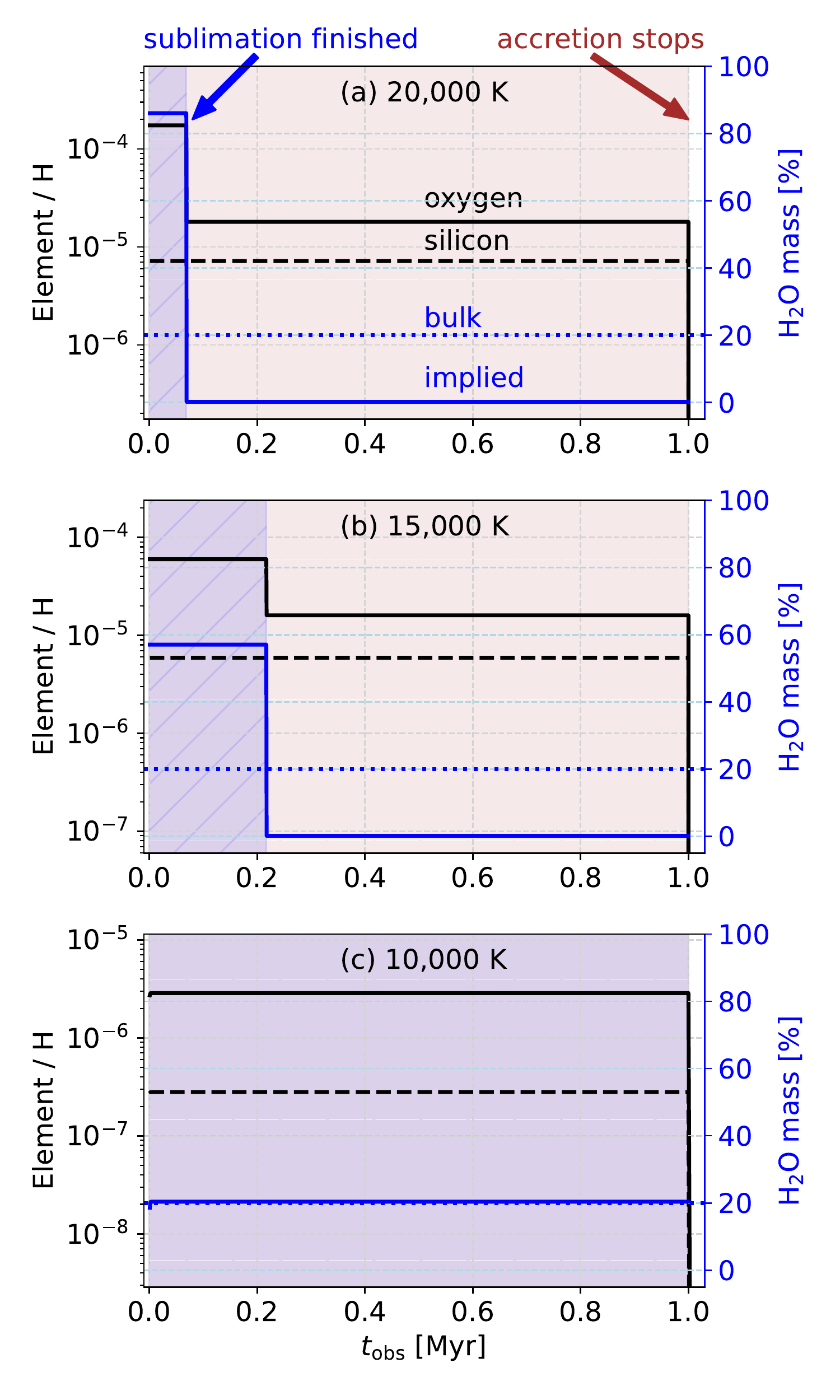}
\caption{Evolution of elemental abundances (black, left axis) and the implied water mass fraction (blue, right axis) when white dwarfs with hydrogen-dominated atmospheres accrete a comet with 20\% ice. In this model, ice accretes at a constant rate until $t_\mathrm{sub}$ (Eq. \ref{eq:t_sub}), and dry rocks accrete in 1 Myr. The water fraction implied by assuming steady state accretion appears discontinuous at $t_\mathrm{sub}$ due to the short diffusion timescales (days here, from \citet{Koester2020} with $\mathrm{log}(g)=8$). The real bulk water content of the object (blue dotted curve) only matches the implied value if $t_\mathrm{sub} >= t_\mathrm{acc}$ yr, as is the case in panel c, at 10,000 K.
\label{fig:evolution_DA}}
\end{figure}
\begin{figure}
\centering
\includegraphics[width=\hsize]{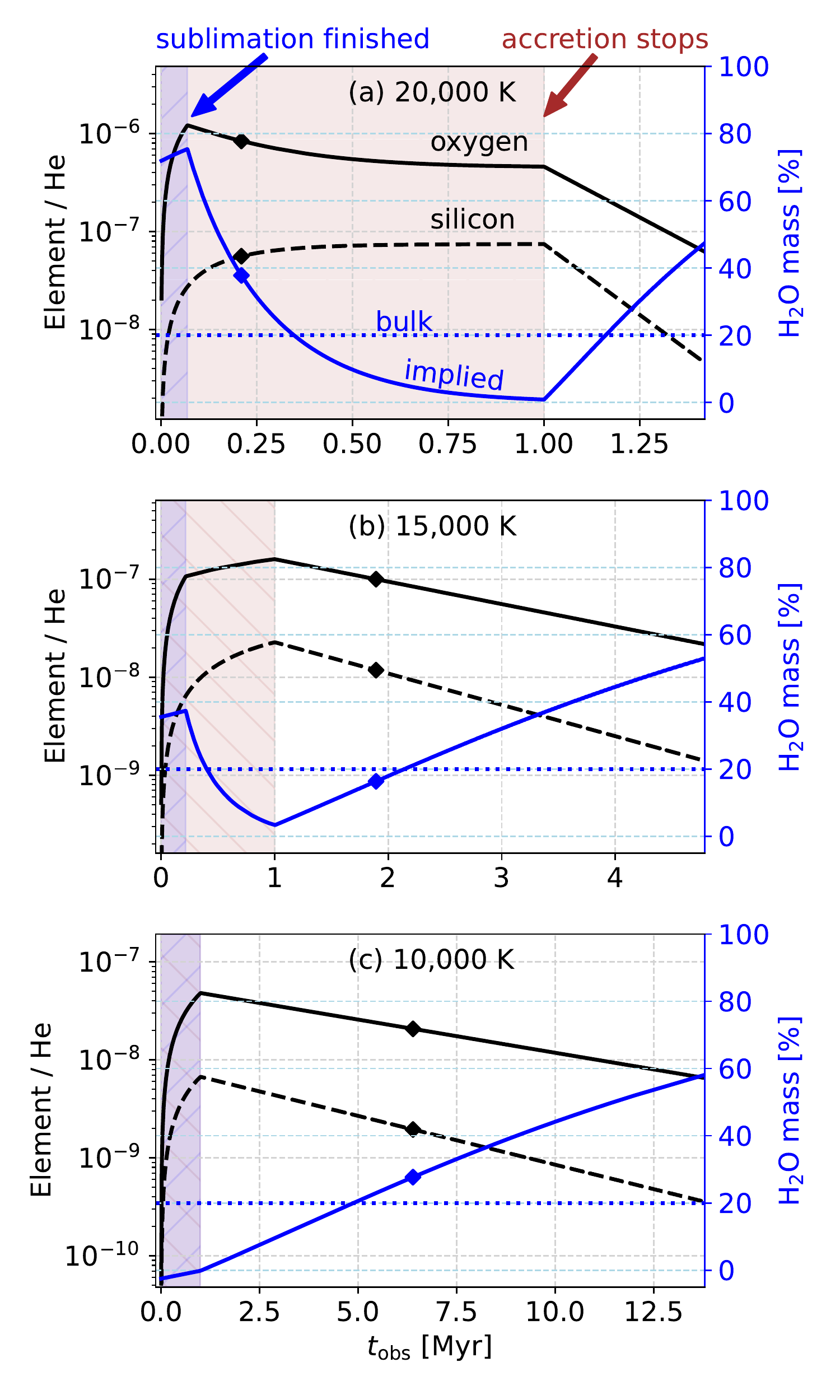}
\caption{Same calculation as shown in Fig. \ref{fig:evolution_DA}, repeated for stars with helium-dominated atmospheres. The squares indicate the sinking timescales of oxygen, while the brown and blue areas indicate the accretion zones of ice and dry rocks, respectively. The implied water content peaks highest for hot stars and is highly dependent on the accretion state, with it increasing further when accretion has stopped and elements sink downwards (declining state).
\label{fig:evolution_DB}}
\end{figure}
We imagine a simple scenario where a comet containing both minerals and ice tidally disrupts around a warm star. Both the rock and ice accretion are illustrative here, and do not include complexities such as the size distribution of fragments and their orbital spread. The rocky material is assumed to accrete evenly over 1 Myr, a typical value implied by comparing DA and non-DA pollution rates \citep{Girven2012, Cunningham2021}. The ice is also modelled to accrete at a constant rate, but with a shorter timescale given by Eq. \ref{eq:t_sub}, assuming a typical fragment size of 1 km. The accretion rates of the different elements onto the star ($\dot{M}_\mathrm{O}, \dot{M}_\mathrm{Si}, ...$) follow from summing over icy and rocky accretion. The rocks have a modified, roughly bulk-Earth elemental composition, set to yield zero oxygen excess \footnote{We take rocks with an Earth-like elemental composition (O:37.3\%, Fe:30.0\%, Si:15.1\%, Mg:14.5\%, Ca:1.6\%, Al:1.5\% by number \citealt{McDonough2003}), slightly rescaled for illustrative purposes to yield zero oxygen excess.}, while the ice is assumed to be pure water ($\mathrm{H_2O}$). The number ratio of an element in the star's convective zone follows from an integration over the accretion rate, modulated by the diffusion timescale ($\tau_\mathrm{O}, \tau_\mathrm{Si}, ...$). This yields the following abundances at time $t_\mathrm{obs}$:
\begin{subequations}
\begin{align}
    \mathrm{O / Hx} (t_\mathrm{obs}) &= \frac{\mu_\mathrm{H_x}}{\mu_\mathrm{O} M_\mathrm{cvz}} \int_{t=0}^{t=t_\mathrm{obs}} \dot{M}_\mathrm{O}(t) \mathlarger{e^{-\left(t_\mathrm{obs}-t\right)/\tau_\mathrm{O}}} dt \label{eq:O_hx} \\
    \mathrm{Si / Hx} (t_\mathrm{obs}) &= \frac{\mu_\mathrm{H}}{\mu_\mathrm{Si} M_\mathrm{cvz}} \int_{t=0}^{t=t_\mathrm{obs}} \dot{M}_\mathrm{Si}(t) \mathlarger{e^{-\left(t_\mathrm{obs}-t\right)/\tau_\mathrm{Si}}} dt, \label{eq:Si_hx} \\
    &\; ... \nonumber
\end{align}
\end{subequations}
where $\mathrm{H_x}$ refers to the dominant atmospheric element, either hydrogen or helium. To infer a pollutant's composition, the atmosphere is often assumed to be in a steady-state between accretion and diffusion. In such a steady-state, the total accretion rate $\dot{M}$ onto the white dwarf is given by:
\begin{equation}\label{eq:Mdot_ss}
    \dot{M} = \sum_{\mathrm{elements} \;  \mathrm{El_i}} \mathrm{El_i}/\mathrm{H_x} \frac{\mu_\mathrm{El_i} M_\mathrm{cvz}}{\mu_\mathrm{H_x}\tau_\mathrm{El_i}},
\end{equation}
where $\mu_\mathrm{i}$ is the atomic weight of element $\mathrm{El_i}$, and $M_\mathrm{cvz}$ is the mass of the star's convective zone. The diffusion timescales and $M_\mathrm{cvz}$ are calculated using the methods described in \citet{Koester2020}\footnote{Extensive tables and more details
of the calculations can be found at \url{http://www1.astrophysik.uni-kiel.de/~koester/astrophysics/astrophysics.html}}, using an overshoot prescription with one pressure scale height of additional mixing. Diffusion timescales vary over many orders of magnitude within the range of effective temperatures from 20,000 to 5,000 K; from days to Myr for hydrogen-dominated atmospheres, and between $10^4-10^8$ yr for helium-rich atmospheres. The oxygen excess associated with accretion refers to the fraction of oxygen atoms that remain after accounting for oxygen bindings in a pre-defined set of minerals. If an element $\mathrm{El_i}$ binds with $n_\mathrm{i}$ oxygen atoms in the minerals of the pollutant ($n_\mathrm{i}=1$ in $\mathrm{FeO, CaO, MgO}$, $n_\mathrm{i}=1.5$ in $\mathrm{Al_2O_3}$, and $n_\mathrm{i}=2$ in $\mathrm{SiO_2}$), the remaining rate of excess oxygen accretion ($\delta \dot{M}_\mathrm{O}$) is:
\begin{equation}\label{eq:delta_Mdot_O}
    \delta \dot{M}_\mathrm{O} =  \mathrm{O}/\mathrm{H_x} \frac{\mu_\mathrm{O} M_\mathrm{cvz}}{\mu_\mathrm{H_x}\tau_\mathrm{O}} - \sum_{\mathrm{elements} \;  \mathrm{El_i}} \mathrm{El_i}/\mathrm{H_x} \frac{n_\mathrm{i} \mu_\mathrm{El_i} M_\mathrm{cvz}}{\mu_\mathrm{H_x}\tau_\mathrm{El_i}}.
\end{equation}
Assuming that these excess oxygen atoms correspond to the accretion of water ice from the pollutant, its implied mass fraction of $f\mathrm{H_2O}$ (in steady-state) follows from:
\begin{equation}
    f\mathrm{H_2O} = \frac{\delta \dot{M}_\mathrm{O} \left(1+2\mu_\mathrm{H}/\mu_\mathrm{O}\right)}{\dot{M}}.
\end{equation}

The evolution of a comet's implied water content during an asynchronous accretion event is illustrated in Figs. \ref{fig:evolution_DA} and \ref{fig:evolution_DB}. For white dwarfs with hydrogen-dominated atmospheres (Fig. \ref{fig:evolution_DA}), which have short diffusion timescales ($\sim$ days), this produces two discontinuous accretion stages, indicated by the blue and brown areas in the figure. In the first stage, the comet's volatiles rapidly sublimate, and the white dwarf accretes material with a water fraction that far exceeds that of the comet. If the atmospheric abundances in this phase are used to infer the cometary composition, its water content would be over-estimated. In contrast, the second accretion phase is entirely dry, and if the star is observed during this period, the pollutant would incorrectly be identified as a dry asteroid. The figure shows that the effects of asynchronous accretion are greater around hotter stars (top panel), which trigger a short and intense phase of sublimative erosion, while the majority of the accretion event is dry.

For stars with helium-dominated atmospheres (Fig. \ref{fig:evolution_DB}), a similar argument applies, but the abundances are more averaged out due to the longer diffusion timescales, which means that their implied oxygen excess is not discontinuous. Instead, the implied oxygen excess peaks at $t_\mathrm{sub}$, and decreases from there. However, with slow downward diffusion, heavy elements continue to be visible in helium-rich atmospheres after all accretion has stopped. During this \textit{declining} phase, the implied oxygen excess generally increases over time, as the oxygen atoms sink more slowly than the other elements, and remain in the atmosphere for longer.

\begin{table*}
\caption{All 9 white dwarfs with previously suggested oxygen excesses. The state-corrected calculation accounts for the likelihood distribution of different accretion states (see Section \ref{sect:oxygen_excess_pylluted}). For systems in blue, we find oxygen excesses with $\geq 2\sigma$ significance with this state-corrected calculation. No new systems with significant oxygen excesses were found.}
\label{table_excesses}
\centering
\addtolength{\tabcolsep}{-0.9 mm}
\begin{tabular}{l c c c c}
\hline
\hline
System & State-corrected excess [\%] & sig. [$\sigma$] & Steady-state excess [\%] & Previous estimate [\%] \\ 
\hline
\sig{\textbf{GALEXJ2339}}${}^\mathrm{(He)}$&61${}^{+8}_{-12}$& 2.82&53${}^{+9}_{-14}$& $\sim 66^{\star,(a)}$\\
\vspace{0.08cm}
\sig{\textbf{WD1425+540}}${}^\mathrm{(He)}$&80${}^{+10}_{-19}$& 2.29&75${}^{+12}_{-23}$& $55^{\star,\bullet,(b)}$\\
\vspace{0.08cm}
\sig{\textbf{WD1232+563}}${}^\mathrm{(He)}$&57${}^{+14}_{-21}$& 2.14&53${}^{+15}_{-21}$& $57^{\star,\ddagger,(c)}$\\
\vspace{0.08cm}
GD61${}^\mathrm{(He)}$&27${}^{+9}_{-12}$& 1.86&27${}^{+8}_{-10}$& $42^{\star,(d)}$, $50^{\dagger,(d)}$\\
\vspace{0.08cm}
GD378${}^\mathrm{(He)}$&73${}^{+15}_{-36}$& 1.49&73${}^{+15}_{-35}$& $\sim 66^{\star,(a)}$\\
\vspace{0.08cm}
WD1536+520${}^\mathrm{(He)}$&20${}^{+28}_{-44}$& 0.49&3${}^{+32}_{-48}$& $<0^{\star,(e)}$, $43^{\dagger,(e)}$\\
\vspace{0.08cm}
SDSSJ1242+5226${}^\mathrm{(He)}$&11${}^{+22}_{-30}$& 0.42&48${}^{+12}_{-16}$& ${57^{+7}_{-7}}^{\star, (f)}$\\
\vspace{0.08cm}
SDSSJ2047-1259${}^\mathrm{(He)}$&1${}^{+27}_{-36}$& 0.02&21${}^{+19}_{-26}$& ${16^{+20}_{-27}}^{\star,(g)}$\\
\vspace{0.08cm}
SDSSJ0956+5912${}^\mathrm{(He)}$&-20${}^{+50}_{-84}$& -&53${}^{+19}_{-34}$& $\sim 45^{\star,(h)}$\\
\hline
\end{tabular}
\begin{tablenotes}
\item ${}^{(a)}$\citet{Klein2021}, ${}^{(b)}$ \citet{Xu2017}, ${}^{(c)}$\citet{Xu2019}, ${}^{(d)}$\citet{Farihi2013},${}^{(e)}$ \citet{Farihi2016}, ${}^{(f)}$\citet{Raddi2015}, ${}^{(g)}$\citet{Hoskin2020}, ${}^{(h)}$\citet{Hollands2022}
\item \textbf{Notes: } $^{(\star)}$ Assuming steady-state accretion. $^{(\dagger)}$ Assuming accretion in a build-up phase (pre steady-state). $^{(\ddagger)}$ Corresponds to a more conservative oxygen assignment with $\mathrm{Fe_2O_3}$ instead of $\mathrm{FeO}$.$^{(\bullet)}$ Corresponds to a differently defined oxygen excess. Re-calculation with our definition yields an excess of 84\%.
\end{tablenotes}
\end{table*}
\section{Observational tests of asynchronous ice-refractory accretion}\label{sect:observational_comparison}
From the preceding arguments, the asynchronous accretion of volatiles and refractories makes two observational predictions:
\begin{enumerate}
    \item There is an anti-correlation between white dwarf temperature and the inferred fraction of wet pollutants. This trend should continue down to $t_\mathrm{acc}=t_\mathrm{sub}$ ($\sim 10,000$ K) for white dwarfs with hydrogen-dominated atmospheres, and down to $t_\mathrm{acc}=t_\mathrm{sink}$ for those with helium-dominated atmospheres ($\sim 15,000$ K).
    \item Some hot white dwarfs will be found with volatile abundances that far exceed the plausible range for comets. These systems can be explained as examples where accretion is caught in the early phase of sublimative erosion.
\end{enumerate}
In this section, we analyse with the current white dwarf sample to compare with these predictions.

\subsection{Description of white dwarf sample}
We compile a new sample of white dwarfs from the literature, which contains all currently identified white dwarfs with a published abundance of atmospheric oxygen, as well as measured abundances of Mg, Si, and Fe. Out of these stars, 9 were previously suggested to have accreted material with an oxygen excess. The complete sample is listed in Appendix \ref{appendix:full_sample}. It is by no means uniformly selected, as oxygen measurements are often performed as a follow-up of the most highly polluted or interesting systems. The majority of the sample has oxygen detections (28/32), with only 4 upper limits. The sample is skewed towards hot white dwarfs (median 14,754 K) with high levels of pollution and helium-dominated atmospheres (23/32), despite the fact that white dwarfs with hydrogen-dominated atmospheres are generally more common.

\subsection{Calculation of oxygen excess with \texttt{PyllutedWD}}\label{sect:oxygen_excess_pylluted}
The amount of oxygen in excess of that required for metal oxides is determined based on the photospheric abundances of all other elements, adjusted to consider different rates of downward diffusion. When observing metals in the atmosphere of a white dwarf, it is not clear a-priori whether accretion is ongoing or finished, with the observed metals caught in the process of sinking out of sight. Whilst clues such as infrared emission, the presence of circumstellar gas or a short sinking timescale (days) might indicate that ongoing accretion is more likely, for most of the white dwarfs considered here, it constitutes a major uncertainty in interpreting the observed abundances.

In order to address this uncertainty here, and to interpolate any missing abundances (e.g. Ca, Al), we make use of the open-source python code \texttt{PyllutedWD}\footnote{\url{https://github.com/andrewmbuchan4/PyllutedWD_Public}} \citep{Harrison2021b, Buchan2022}. \texttt{PyllutedWD} works from the principle that a range of accretion states might be consistent with the data within the error range, but that the observed abundances can provide clues as to the most likely values of the current accretion state. In this physical model, elements with different condensation temperatures (e.g. Ca relative to Mg or Na) or different affinities to enter the iron melt (e.g. Ca/Fe or Ni/Fe) can vary in ways appropriate to volatile loss and core-mantle differentiation. The only other process in this model that can alter key ratios of elemental pairs with similar condensation temperatures is relative sinking due to variations in the rate of downward diffusion. By considering only these processes, insights regarding the most likely accretion state of the white dwarf pollutants can be determined, which for some individual systems constrain the accretion to be most likely in steady-state (or declining), whilst for others systems, a broad likelihood distribution of the accretion states can be obtained.

To calculate the oxygen excess of a pollutant, we go through the following procedure: First, we run \texttt{PyllutedWD} to find a posterior distribution of the accretion state (see Fig. \ref{fig:accretion_posterior}). Next, to account for the errors, we sample the pollutant's abundances $10^4$ times from Gaussian functions centred on the reported or inferred values, taking a conservative error of 0.4 dex on any inferred minor abundances (Al, Ca). For each sampling, an accretion state is assigned by a weighted draw from the posterior distribution of accretion states, which is used to translate the photospheric abundances into pollutant abundances. Generalizing from Eqs. \ref{eq:O_hx} and \ref{eq:Si_hx}, the accreted abundance ratio $\mathrm{El_1/El_2}$ for elements 1 and 2 that accrete at a constant rate from $t=0$ to $t=\mathrm{min}(t_\mathrm{acc}, t_\mathrm{obs})$ is:
\begin{equation}
    \frac{\mathrm{El_1}}{\mathrm{El_2}} = \frac{\mathrm{El}_1(t_\mathrm{obs})}{\mathrm{El}_2(t_\mathrm{obs})} \frac{\tau_\mathrm{El_2}}{\tau_\mathrm{El_1}}
    \mathlarger{
    e^{\left(\mathrm{min}(t_\mathrm{acc}, t_\mathrm{obs})-t_\mathrm{obs}\right)\left(1/\tau_\mathrm{El_2}-1/\tau_\mathrm{El_1}\right)}
    }.
\end{equation}
Calculated for the complete posterior distribution of accretion states and abundance samplings, this produces $10^4$ sets of equally likely abundances for each pollutant, adjusted for relative diffusion rates. Next, we follow the standard procedure outlined by \citet{Klein2010}, where the major rock-forming elements are assigned to the minerals $\mathrm{FeO, CaO, SiO_2, Al_2O_3,MgO}$. A portion of the iron can also be present in metallic form, but this fraction is unknown. From the oxygen assignments to minerals, we obtain a histogram of oxygen excesses for each pollutant, expressed as a number fraction of the total oxygen abundance (see Fig. \ref{fig:oxygen_excess_posterior}). The oxygen excess is given as the median value of this distribution, with asymmetric $1\; \sigma$ errors identified from the $15.8^\mathrm{th}$ and $84.1^\mathrm{st}$ percentiles. The total significance is given by the fraction of this distribution with a positive oxygen excess. Our statistical approach yields wide errors on the oxygen excesses of most systems, due to a combination of the uncertainties on abundances and accretion states. Nevertheless, we note that the true uncertainties on the calculated excesses are likely to exceed our quoted values, mainly due to additional expected inaccuracies in relative diffusion timescales (see \citealt{Blouin2020, Heinonen2020}) that are not accounted for in this analysis. Further challenges are discussed in Section \ref{sect:discussion_challenges}

\subsection{Prediction I: Oxygen excess and stellar temperature}
\begin{figure}
\centering
\includegraphics[width=\hsize]{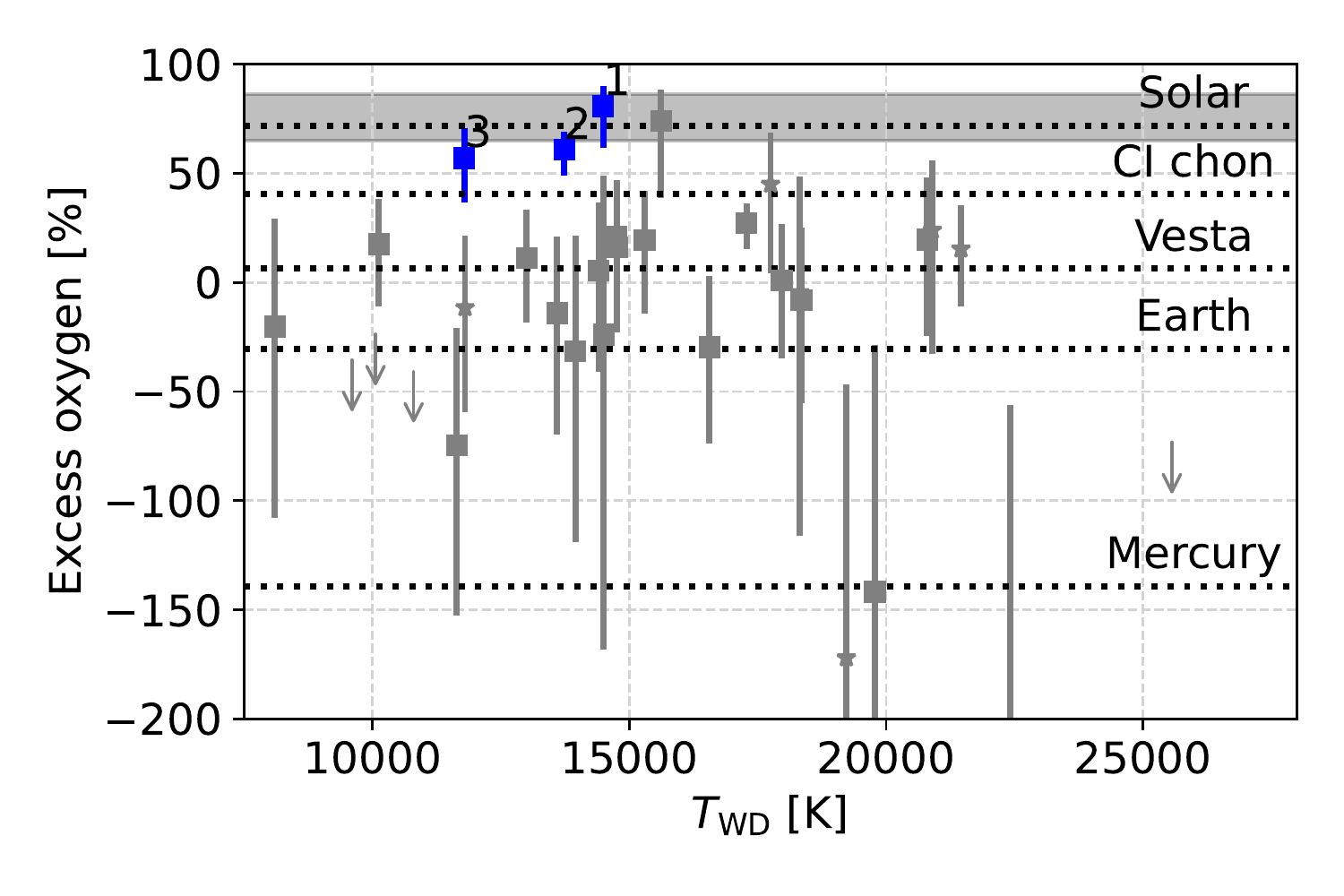} 
\caption{Oxygen excesses of white dwarfs in our sample (star; hydrogen-dominated, square; helium-dominated), compared with Solar System bodies (Mercury: \citealt{Hauck2013, Nittler2018}, Earth: \citealt{McDonough2003}, Vesta: \citealt{Steenstra2016}, CI chondrites \& Solar photosphere \citealt{Lodders2003}, FGK stars (grey band; \citealt{Brewer2016})). The three blue points signal white dwarfs with significant ($\geq 2 \sigma$) oxygen excesses in the default, state-corrected calculation (see Section \ref{sect:oxygen_excess_pylluted}). These labelled systems are 1: WD1425+540, 2: GALEXJ2339, and 3: WD1232+563. \label{fig:exO_trends}}
\end{figure}

The first prediction from asynchronous ice-refractory accretion is that there exists an anti-correlation between white dwarf temperature and the inferred fraction of wet pollutants. In Fig. \ref{fig:exO_trends}, the oxygen excesses of all white dwarfs in the sample are plotted against stellar temperature. We find that almost all systems are consistent with the accretion of dry rocks, with only three systems showing an oxygen excess at the $\geq 2\sigma$ level (shown in blue). This threshold is set at $2\sigma$ for practical reasons, as a lower level of $1\sigma$ yields too many expected false positives (4.4, against 0.62), and the uncertainties on abundances and accretion states are currently too great to reach $3\sigma$ for any system. The limited number of systems with oxygen excesses at even $2\sigma$ significance highlights that the uncertainties on the accretion states and relative abundances make it difficult to ascertain oxygen excesses with strong statistical certainty. In addition, we note that no systems in the sample showed evidence of being significantly reduced in oxygen, consistent with prior analysis by \citet{Doyle2020}.

The three white dwarfs with oxygen-rich pollutants all have effective temperatures at or below the sample median of 14,500 K. While this can be interpreted as tentative evidence for the proposed anti-correlation between oxygen excesses and white dwarf temperature, the number of systems is currently too small for a statistical argument to be made. Furthermore, the three white dwarfs with significant oxygen excesses all have helium-dominated atmospheres with long diffusion timescales, making their accretion states subject to greater uncertainty. In the future, when oxygen lines are identified in a larger number of white dwarfs with hydrogen-dominated atmospheres, and abundance errors are reduced, this new sample would be ideally suited to investigate the proposed anti-correlation between white dwarf temperature and the volatile content in their accretion.

\subsection{Prediction II: rare instances of near-pure ice accretion}
\begin{figure}
\centering
\includegraphics[width=\hsize]{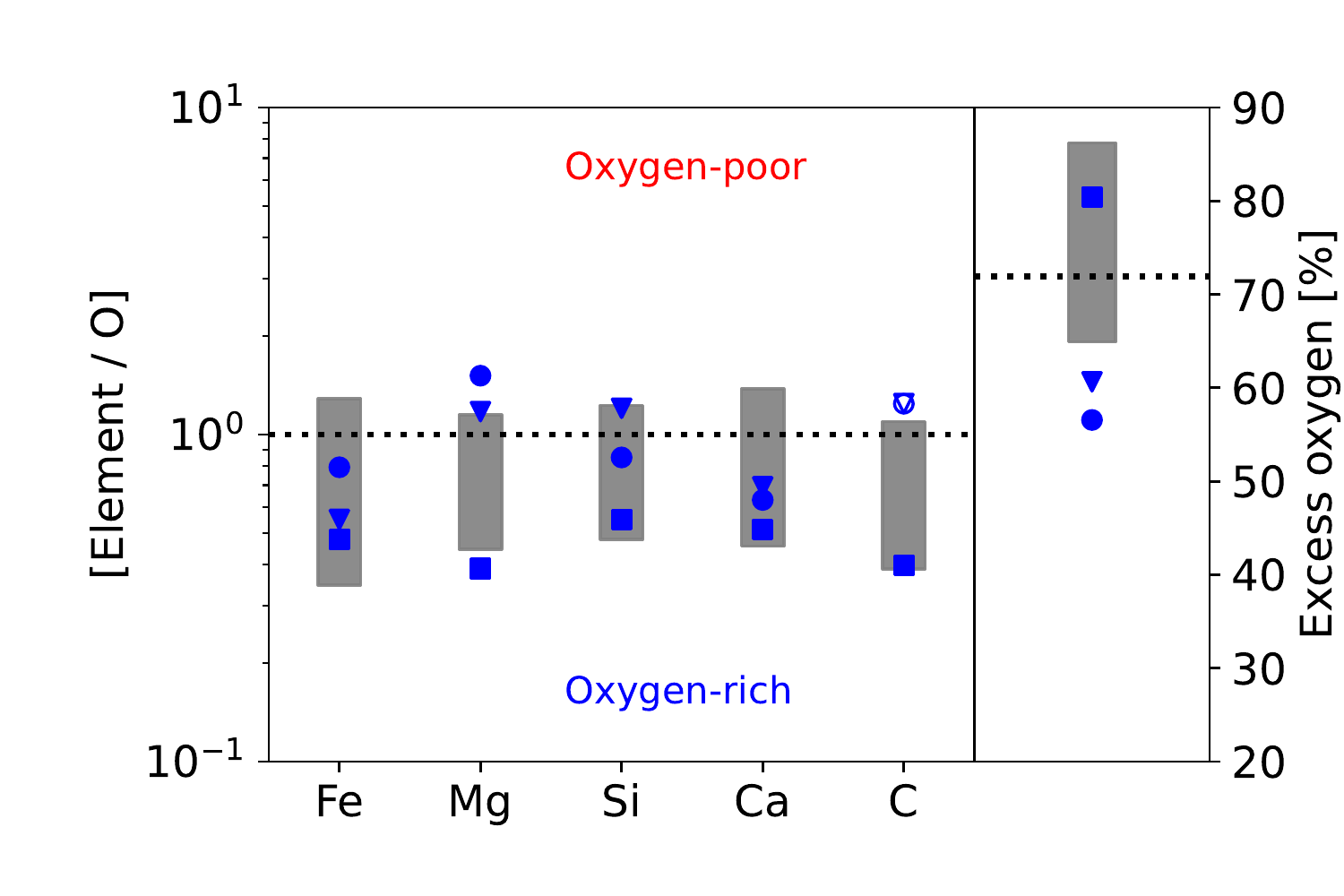}
\caption{Comparison of oxygen-rich pollutants ($\geq 2\sigma$, blue) to a sample of nearby FGK stars (grey band, \citealt{Brewer2016}), plotted relative to oxygen, relative to the Solar ratio. Filled markers represent measured abundances (square; WD1425+540, triangle; GALEXJ2339, circle; WD1232+563), while open markers are any non-measured abundances, inferred by \texttt{PyllutedWD}.
\label{fig:comparison_stars}}
\end{figure}
The theoretical `smoking gun` for asynchronous ice-refractory accretion is a hot white dwarf that appears to be accreting material with an almost entirely icy composition (see Figs. \ref{fig:evolution_DA},\ref{fig:evolution_DB}).
Most comets are a mixture of refractories and ice, with the maximum volatile content occurring when material of pure solar (stellar) composition condenses to form dust and ices. If some white dwarfs are accreting material with relative oxygen abundances exceeding stellar values, this is expected by this model as a result of the sequential accretion of volatiles and refractories. To see if any of the pollutants in the sample match near-pure ice accretion, we compare the abundances of the oxygen-rich systems with nearby FGK stars in Fig. \ref{fig:comparison_stars}. The FGK sample was compiled by \citet{Brewer2016} and represents a plausible compositional range for pristine comets with complete condensation. 

Out of the three white dwarfs with significant oxygen excesses, one system (WD1425+540) is found to be highly oxygen-rich. When compared to Solar elemental ratios, the composition of WD1425+540's pollutant is enhanced in oxygen relative to the other major rock-forming elements, but it still falls within the plausible range of FGK stars. If these pollutant abundances represent the bulk content of a comet, it must have undergone minimal thermal processing during the star's post-main sequence evolution. In the models of \citet{Malamud2017a}, complete water retention is only expected if the pollutant originates beyond $\sim$100 AU, which would require it to be analogous to an Oort cloud comet. If the comet originates from a closer orbit, with less ice, asynchronous accretion could explain the high level of oxygen excess, as it is just within the expected temperature range for asynchronous accretion to be noticeable. In the near future, the discovery and analysis of hot white dwarfs with hydrogen-dominated atmospheres will provide opportunities to find more clear-cut examples of asynchronous ice-refractory accretion. We discuss the system SDSS J0914+1914 as an interesting case study in Section \ref{sect:WDJ0914+1914}.

\subsection{Comparison with previous estimates of oxygen excesses}
Our procedure for calculating the oxygen excesses of white dwarfs is generally more conservative than that in previous works. For just three systems with previously suggested oxygen excesses, we find an excess with $\geq 2\sigma$ significance. The inferred oxygen excesses for GALEXJ2339 and WD1225+540 are nearly identical to those reported by \citet{Klein2021} and \citet{Xu2019}. For WD1425+540, we find a substantially higher oxygen excess than reported by \citet{Xu2017}, but this is due to a different definition of the oxygen excess in their work. When the same definition is used, their methodology yields an oxygen excess of 84\% in steady-state, which is in line with our new analysis.

For an additional five systems, our analysis infers a potential oxygen excess, but only at $<2\sigma$. These white dwarfs are unlikely to be accreting dry material, but there is insufficient evidence for wet accretion at this statistical level. The case of GD61 is right on this threshold, with an inferred significance of $1.9\sigma$. For this system, we identify a lower median excess than reported by \citet{Farihi2013} due to longer relative diffusion timescales of oxygen in the updated models of \citet{Koester2020}. In addition, we find a $\sim25 \%$ probability of GD61 being in a declining state. This option was ruled out by \citet{Farihi2013} based on the presence of circumstellar material \citep{Farihi2011}, although a declining state can be consistent with an infrared excess if the system is just accreting at a lower rate than before. Together with the abundance uncertainties, this pushes the significance of the oxygen excess in our computation just below $2\sigma$. A similar argument applies to our results for GD378 \citep{Klein2021}, SDSSJ1242+5226 \citep{Raddi2015}, SDSSJ2047-1259 \citep{Hoskin2020}, and WD1536+520 \citep{Farihi2016}. If the accretion state is not constrained with a very high degree of certainty, as is seldom the case for white dwarfs with helium-dominated atmospheres, the possibility of a declining state will often suppress the significance of an oxygen excess below $2\sigma$.

Finally, our inferred oxygen excess for SDSSJ0956+5912 ($-21{}^{+51}_{-87}\%$, taking the abundances from the Gran Telescopio Canarias reported by \citet{Hollands2022}) differs substantially from the estimates by \citet{Hollands2022}, who reported a value of $\sim45\%$ with the same abundances. In this case, our Bayesian analysis with \texttt{PyllutedWD} confidently infers the accretion to be in a late declining state, at more than two diffusion timescales after accretion. However, \citet{Blouin2020} showed that with the updated constitutive physics in their model, the diffusion timescale of magnesium in old white dwarfs with helium-dominated atmospheres could be shorter than currently modelled, potentially biasing the analysis of this system to a declining state. We further discuss the challenges of confidently inferring accretion states in Section \ref{sect:discussion_challenges}.

\section{Discussion}\label{sect:discussion}

\subsection{Interpretation of dry/wet photospheric abundances}
The predicted rapid sublimative erosion of ices around warm ($\gtrsim 10,000$ K) white dwarfs implies that the material accreting from a single comet may initially appear highly volatile-rich, and become dry as accretion continues. In light of this possibility, we advise a cautious approach to the interpretation of volatile abundances of warm white dwarfs. While the detection of an oxygen excess always points to the presence of ices, the inferred quantity might not match the pollutant's bulk composition. Additionally, abundances that seem to suggest the accretion of a dry asteroid, might in fact correspond to the late accretion stage of a wet comet. The effects of asynchronous ice-refractory accretion are most important for the hottest white dwarfs, especially for those with hydrogen-dominated atmospheres, where downward diffusion of heavy elements does not average over different accretion stages. For the hottest stars, the initial wet accretion phase is expected to be short, with much longer-lasting dry accretion. At no point during the accretion process is it guaranteed that the abundances of a white dwarf atmosphere match the bulk composition of an accreting comet. For white dwarfs with helium-dominated atmospheres, the same argument applies, but only down to $\sim 15,000$ K, when the timescale for the downward diffusion of heavy elements exceeds 1 Myr, and the wet and dry accretion stages likely become indistinguishable.

\subsection{Observational tests for asynchronous volatile-refractory accretion}
The proposed scenario of asynchronous ice-rock accretion makes two observational predictions, and can either be corroborated or ruled out by future data. The first prediction is that there exists an anti-correlation between white dwarf temperature and the inferred fraction of wet pollutants. While we find that there are currently not enough white dwarfs with determined oxygen abundances to investigate this trend, upcoming large-scale spectroscopic surveys (4MOST/WEAVE/DESI/SDSS-V) will provide an opportunity to significantly expand the size of this sample in a uniform manner. White dwarfs with hydrogen-dominated envelopes are most useful for this purpose, as they are almost certainly in a steady-state accretion phase, which is typically necessary for the oxygen excess to be well-constrained. However, we also emphasize the requirement for sufficiently small errors on atmospheric abundances of O, Si, Mg, and Fe, as current limits are insufficient for significant statistical results. A second difficulty for the verification of this prediction is the same trend has previously been predicted based on the hypothesized inside-out scattering of planetesimals from the surrounding planetary system \citep{Jura2014, Malamud2016}. Due to this similar prediction from dynamic models, the detection of such a trend will never provide definite evidence in favour of asynchronous accretion. Recent evidence seems to argue against the inside-out depopulation of planetesimals, however, as the implied decline of the inward scattering rate over time \citep{Li2022} has not been observed \citet{Blouin2022}.

\subsubsection{Pure ice accretion onto SDSS J0914+1914?}\label{sect:WDJ0914+1914}
Secondly, the process of asynchronous ice-rock accretion could be proven by the identification of hot white dwarfs with super-Solar volatile abundance ratios. Out of the three systems in our white dwarf sample identified to be oxygen-rich, none clearly falls within this category. However, there is one example of a white dwarf, SDSS J091405.30+191412.2 \citep{Gaensicke2019}, where only oxygen and sulphur were detected, and so virtually all the accreting oxygen appears to be in the form of ices. Interestingly, SDSS J0914+1914 is extremely warm for a polluted white dwarf ($T_\mathrm{eff}=27,750 \pm 310$ K) and has a hydrogen-dominated atmosphere. At such a high temperature, the sublimative erosion and accretion of ices is expected to dramatically outpace the accretion of minerals, and the short diffusion timescale allows for a clear compositional separation of wet and dry accretion phases (See Fig. \ref{fig:evolution_DA}).

In the work by \citet{Gaensicke2019}, the extreme inferred composition of the accreting material was suggested to match the deep layers of an evaporating ice giant planet. In this scenario, EUV photons from the star ionize hydrogen from the planetary envelope \citep{Bourrier2013, Owen2019}, triggering a hydrodynamic outflow of hydrogen that takes with it significant amounts of $\mathrm{H_2O}$ and $\mathrm{H_2S}$. To distinguish both scenarios, the compositional differences of ice giants and comets can be compared. Most strikingly, only ice giants are massive enough to bind substantial amounts of gaseous hydrogen and helium from their natal environments \citep{Mizuno1980, Bodenheimer1986, Pollack1996}. In the cases of Uranus and Neptune, their bulk gaseous hydrogen content has been modelled at 10\% and 5\%, respectively \citep{Podolak1995}, although both numbers are subject to uncertainty \citep{Helled2020, Podolak2022, Vazan2022}. Compared to these estimates, the disc around SDSS J0914+1914 is depleted in hydrogen by more than an order of magnitude, explained by \citet{Gaensicke2019} as the result of radiation pressure. In Appendix \ref{appendix:rad_pressure}, we verify that radiation pressure can indeed push out neutral hydrogen atoms, but we show that the other rock-forming elements are equally affected. In addition, the pressure felt by hydrogen atoms is strongly dependent on the degree of red/blue shifting, complicating the picture significantly. 

Helium atoms are unaffected by radiation pressure, however, and their abundance is more suitable to distinguish the two accretion scenarios. The chemical inertness of helium makes it extremely depleted in comets, at least by a factor $10^4$ in comet Austin \citep{Stern1992}. In contrast, helium is abundant in all layers of ice giants, which are too small for helium to phase-separate \citep{Mankovich2020, Miguel2022, Guillot2022}, predicting an approximately solar ratio of He/H, similar to that observed in their outer layers \citep{Conrath1987,Conrath1991}. Following the model by \citet{Moses2020}, the ratio of $\mathrm{log(He/O)}$ in Uranus and Neptune drops from virtually zero in their outer layers to 0.5 and -0.22 in their deep interiors, respectively. The current upper limit for SDSS J0914+1914 was determined by \citet{Gaensicke2019} at $\mathrm{log(He/O)}<1.05$, which is not strict enough to distinguish between both scenarios. In the future, a detection of helium at or near the current upper limit can confirm an ice giant scenario, while a tighter constraint on the star's helium abundance below $\mathrm{log(He/O)}\lesssim -1$ would rule it out, favouring a scenario of asynchronous accretion of volatiles and refractories.

\subsection{Uncertainties in volatile/refractory accretion timescales}
The importance of asynchronous volatile-refractory accretion can be quantified by the accretion timescale ratio of ices relative to rocks. Currently, both timescales are subject to significant uncertainties. In our calculation of sublimative erosion, we assume that the sublimation occurs in a highly eccentric disc following a tidal disruption. If the accretion process occurs in a tighter, circular disc, the sublimation timescale becomes substantially shorter. Secondly, the interiors of cometary fragments likely consist of a complex matrix of different minerals and ices, with some porosity depending on its interior pressure (\citealt{Leliwa2000, Durham2005, Yasui2009}), rather than a single species. When a fragment is insufficiently permeable, vapour could remain trapped inside, shielded by an outer dust layer, making volatile escape take a different form \citep{Fulle2019, Fulle2020, Gundlach2020}. Modelling of such processes remains contingent on various uncertain thermal and permeability coefficients, even in current state-of-the-art models \citep{Davidsson2021,Malamud2022}. Additionally, in the geometry of an eccentric tidal disc, the environment can become radially opaque, in which case the radiative flux received by fragments depends largely on the vertical structure of the disc. Finally, a portion of a comet's volatiles may also be lost prior to its tidal disruption, during the preceding tidal evolution (\cite{Malamud2016, Malamud2017a, Malamud2017b}, Pham et al, in prep.). Such a scenario of asynchronous accretion can also lead to pure volatile accretion, albeit at lower accretion rates.

On the refractory side of the comparison, the accretion timescale of rocky material is also not very well constrained. It was inferred to be between $10^4-10^6$ yr by \citet{Girven2012}, based on the difference in pollution rates between white dwarfs with hydrogen and helium-dominated atmospheres. This estimate was later adjusted upwards to $10^5-10^7$ yr by \citet{Cunningham2021} using updated diffusion timescales. However, bursts of rapid accretion of rocky material might be required to explain the occasional measurement of extremely highly polluted atmospheres \citep{Farihi2012b}. If rocky accretion timescales are indeed on this shorter side of the estimations, asynchronous ice-rock accretion only operates if tidal disruptions produce icy fragments smaller than 100 m, which could still be reasonable given the relatively weak material strength of ice \citep{Greenberg1995, Davidsson1999, Gundlach2016}. 

\subsubsection{Vapour accretion after sublimative erosion}
Besides uncertainties relating to sublimative erosion itself, the accretion of eccentric gas onto white dwarfs has not been studied in detail. \citet{Trevascus2021} showed that gas released by a body on a moderately eccentric trajectory can remain on a similar orbit for several orbital periods due to aerodynamic coupling, but it is not clear if this generalizes to the highly eccentric ($e\sim0.999$) orbits expected after a tidal disruption. It is unlikely that the vapour promptly circularizes via shocks induced by orbit crossings analogous to stellar disruption around black holes \citep{Rees1988}, as the gravitational field is much weaker than this analogous case. When circularization does occur, energy released by the process will heat the gas to a highly ionized state. At this point, its effective viscosity is likely increased by the magneto-rotational instability, and most of the gas can viscously accrete within tens of years \citep{Rafikov2011b}.

\subsection{Challenges in accurately constraining the accretion state}\label{sect:discussion_challenges}
To confidently infer the oxygen excess of pollutants, or more broadly their composition, the accretion state of the system must be well-constrained. In this work, we use the Bayesian code \texttt{PyllutedWD} \citep{Buchan2022} for this purpose (see Fig. \ref{fig:accretion_posterior} for the posterior distributions). Our approach has the advantage of yielding a likelihood distribution of possible accretion states, constrained by all available elements, but is still limited in three ways. First, it assumes a physical model for the formation of a pollutant (described by \citealt{Harrison2021b}, see also \citealt{Rogers2022}) that could be over-simplified. Second, if the true diffusion timescales of particular elements differ from the modelled values (in our case by \citealt{Koester2020}), the Bayesian model will be biased towards a particular accretion state. The uncertainty of accretion timescales is a known issue for cool white dwarfs with helium-dominated atmospheres \citep{Blouin2020, Heinonen2020}. Similarly, because the errors on the diffusion timescales of elements are due to systematics, and therefore largely unknown, the true distribution of consistent accretion states is likely to be broader than our modelled values. 

More generally, because the accretion states of white dwarfs with helium-dominated atmospheres are difficult to constrain in practice, accounting for the relevant uncertainties almost always produces a broad range of possible pollutant compositions. In our analysis, this means that no systems can currently be claimed to have swallowed oxygen-rich pollutants with $3\sigma$ significance. For only three systems, the abundances are such that we infer an oxygen excess at $\geq 2\sigma$. To improve the constraints on accretion states, it helps if the abundances and upper limits of as many elements as possible are reported (see \citealt{Harrison2021b, Buchan2022}), but significant uncertainties are likely to remain. In light of this, it might be that white dwarfs with hydrogen-dominated atmospheres are most suitable for statistically strong claims about the compositions of their pollutants, and that future work should focus on building a sizeable sample of polluted white dwarfs with this atmospheric type. However, in the current sample, the abundance errors on the white dwarfs with hydrogen-dominated atmospheres are also prohibitively large for strong statistical statements about their water content, showing that reduced uncertainties are required across the board to allow the presence of water to be statistically proven.

\section{Summary and conclusions}\label{sect:conclusions}
The presence of water ice in planetesimals may be necessary to form habitable planets in the otherwise dry inner zones of planetary systems. Polluted white dwarfs provide a unique opportunity to assess this water content in systems around other stars \citep{Farihi2013, Farihi2016, Raddi2015, Xu2017, Xu2019, Hoskin2020, Klein2021, Hollands2022}, but the analysis is made complex by details of the accretion process. In this work, we study the pollution of white dwarfs with cometary material, and investigate the scenario suggested by \citet{Malamud2016}, where icy fragments potentially sublimate and accrete prior to the pollutant's rocky components. We find that the timescale for sublimative erosion is indeed short, less than 1 Myr when the stellar temperature exceeds $10,000$ K (Eq. \ref{eq:t_sub}). Based on this result, we suggest that the accretion of a single comet might begin with a volatile-rich phase, followed by the accretion of dry dust (Figs. \ref{fig:evolution_DA}, \ref{fig:evolution_DB}). The proposed scenario of asynchronous ice-rock accretion makes two testable predictions:
\begin{enumerate}
    \item There is an anti-correlation between white dwarf temperature and the inferred fraction of wet pollutants. This trend should continue down to $t_\mathrm{acc}=t_\mathrm{sub}$ ($\sim 10,000$ K) for white dwarfs with hydrogen-dominated atmospheres, and down to $t_\mathrm{acc}=t_\mathrm{sink}$ for those with helium-dominated atmospheres ($\sim 15,000$ K).
    \item Some hot white dwarfs will be found with volatile abundances that far exceed the plausible range for comets. These systems can be explained as examples where accretion is caught in the early phase of sublimative erosion.
\end{enumerate}
To test these predictions, we collate and analyse a sample of white dwarfs from the literature. We find that due to large combined uncertainties on abundances and accretion states, only three systems show significant ($2\sigma$) evidence of water, an insufficient number to investigate the first prediction. Incidentally, the scenario of asynchronous accretion might explain the extremely volatile composition of material accreting onto SDSS J0914+1914, where only oxygen and sulphur were observed. To distinguish this scenario from the accretion of an ice giant, proposed by \citet{Gaensicke2019}, a stricter constraint on the helium abundance is required. 

Given our results, we advise a cautious approach to the interpretation of volatile abundances of warm ($>10,000$ K) white dwarfs. While the identification of an oxygen excess points to the presence of water in the accreting material, the inferred value does not necessarily match the bulk composition of the comet. Similarly, a pollutant that appears to be dry, might in fact be volatile-rich if it is observed in the later stages of its accretion. Finally, we show the importance of interpreting pollutant compositions in a statistical manner, and highlight the difficulty of making statistically strong statements given the large uncertainties on abundances and accretion states. For stars with helium-dominated atmospheres, accounting for uncertainties in their accretion state alone (build-up/steady-state/declining) often already produces a broad range of possible pollutant compositions. In this work, uncertainties on abundances and accretion states limit the significance of most inferred oxygen excesses below $2\sigma$. In light of this, we emphasize the need for reduced uncertainties on atmospheric abundances, and encourage observational efforts to focus on expanding the sample of polluted white dwarfs with hydrogen-dominated atmospheres, whose accretion states are typically well-constrained.

\section*{Data availability}
The simulation data that support the findings of this study are available upon request from the corresponding author, Marc G. Brouwers.

\section*{Acknowledgements}
We thank the anonymous referee for comments that improved the clarity and presentation of our work. Marc G. Brouwers acknowledges the support of a Royal Society Studentship, RG 16050. Andrew M. Buchan is grateful for the support of a PhD studentship funded by a Royal Society Enhancement Award, RGFEA180174. Amy Bonsor acknowledges support from a Royal Society Dorothy Hodgkin Research Fellowship (grant number DH150130) and a Royal Society University Research Fellowship (grant number URFR1211421). Elliot M. Lynch was supported by the ERC through the CoG project PODCAST No 864965 and the European Union’s Horizon 2020 research and innovation program under the Marie Sklodowska-Curie grant agreement No 823823. Laura K. Rogers is grateful for PhD funding from STFC and the Institute of Astronomy, University of Cambridge.

\bibliographystyle{mnras}
\bibliography{circularisation}

\begin{appendix}
\section{Neglection of heat diffusion during fragment sublimation}\label{appendix:conduction}
In this appendix, we validate the assumption in Section \ref{sect:sublimation} that inward heat transport can be neglected in the energy balance of sublimating icy fragments around a white dwarf. The key variable to calculate is the time ($\delta t$) that a fragment spends inside the sublimation zone where $T>T_\mathrm{sub}$ around the star. We begin with an estimation of the distance ($d_\mathrm{sub}$) for the onset of sublimation, which follows from the equilibrium temperature:
\begin{equation}
    d_\mathrm{sub} = \frac{R_\mathrm{WD}\sqrt{1-A}}{2} \left(\frac{T_\mathrm{WD}}{T_\mathrm{sub}}\right)^2.
\end{equation}
The edge of the sublimation zone has the following eccentric anomaly ($E_\mathrm{sub}$) and mean anomaly ($M_\mathrm{sub}$):
\begin{subequations}
\begin{align}
    1-e\; \mathrm{cos}\left(E_\mathrm{sub}\right) &= d_\mathrm{sub}/a \label{eq:Esub}\\
    M_\mathrm{sub} &= E_\mathrm{sub}- e\;\mathrm{sin}\left(E_\mathrm{sub}\right). \label{eq:Msub}
\end{align}
\end{subequations}
The time spent in the sublimation zone is given by $\delta t = 2M_\mathrm{sub}/\bar{\omega}$ with $\bar{\omega}=\sqrt{G M_\mathrm{WD}/a^3}$. The equations above do not have a general closed form solution, and we instead perform an asymptotic expansion, using the following substitutions:
\begin{equation}
    e=1-\epsilon\;,\; E_\mathrm{sub}=\delta^{1/2} x_\mathrm{sub} \;,\; \lambda = \delta/\epsilon,
\end{equation}
where $\epsilon, \delta$ are both small numbers and $x_\mathrm{sub}, \lambda$ are order unity. 
With these substitutions, Eqs. \ref{eq:Esub} and \ref{eq:Msub} reduce to the following leading order terms in $\epsilon$:
\begin{subequations}
\begin{align}
    x_\mathrm{sub} &= \sqrt{\frac{2}{\lambda}} \left(\frac{d_\mathrm{sub}}{a\epsilon}-1\right)^{1/2}, \\
    M_\mathrm{sub} &= \epsilon^{3/2} \lambda^{1/2} x_\mathrm{sub} \left(1+\frac{\lambda x_\mathrm{sub}^2}{6}\right).
\end{align}
\end{subequations}
Together, the factors $\lambda$ cancel out, and we find the time spent in the sublimation zone:
\begin{align}
    \delta t_\mathrm{sub} &= \frac{\sqrt{2}\left(1-e\right)^{3/2}}{3\bar{\omega}} \sqrt{\frac{d_\mathrm{sub}}{r_\mathrm{B}}-1} \left(\frac{d_\mathrm{sub}}{r_\mathrm{B}}+2\right) \label{eq:delta_t} \\
     &= \left\{\begin{array}{l}
     \frac{\sqrt{2}\left(1-e\right)^{3/2}}{\bar{\omega}} \left(\frac{d_\mathrm{sub}}{r_\mathrm{B}}-1\right)^{1/2} \qquad \mathrm{if \; d_\mathrm{sub} \sim r_\mathrm{B}}
     \\
     \frac{\sqrt{2}}{3\bar{\omega}} \left(\frac{d_\mathrm{sub}}{a}\right)^\frac{3}{2}
     \quad\qquad\qquad\qquad \mathrm{if \;  d_\mathrm{sub} \gg r_\mathrm{B}}, \end{array}\right.
\end{align}
where $r_\mathrm{B}=a(1-e)$. The time spent inside the sublimation zone does not depend on the semi-major axis of the fragment, which cancels out in Eq. \ref{eq:delta_t}.

During the time $\delta t_\mathrm{sub}$ that the fragment spends inside the sublimation zone, a portion of the heat will conduct inward and penetrate to deeper layers, as described by the spherical heat diffusion equation. A typical estimate of this distance is defined by the so-called skin depth \citep[e.g.,][]{Huebner2006}:
\begin{equation}
    \delta R_\mathrm{skin} =  \sqrt{\frac{K_\mathrm{th} \delta t_\mathrm{sub}}{\pi \rho C_\mathrm{p}}}.
\end{equation}
In the most interesting scenario of rapid sublimation ($d_\mathrm{sub} \gg r_\mathrm{B}$), the skin depth during sublimation is:
\begin{align}
    \delta R_\mathrm{skin} &\simeq 6 \; \mathrm{cm} \left(\frac{T_\mathrm{WD}}{10^4 \; \mathrm{K}}\right)^{\frac{3}{2}} \left(\frac{T_\mathrm{sub}}{300 \; \mathrm{K}}\right)^{-\frac{3}{2}} \left(\frac{R_\mathrm{WD}}{\mathrm{R_\oplus}}\right)^{\frac{3}{4}}
    \left(\frac{M_\mathrm{WD}}{0.6\;\mathrm{M_\odot}}\right)^{-\frac{1}{4}} \label{eq:dR_skin} \\
    &
    \left(\frac{\rho_\mathrm{frag}}{1 \; \mathrm{g/cm^3}}\right)^{-\frac{1}{2}}
    \left(\frac{C_\mathrm{p}}{2 \cdot10^7\; [\mathrm{cgs}]}\right)^{-\frac{1}{2}} \left(\frac{K_\mathrm{th}}{2\cdot10^5\; [\mathrm{cgs}]}\right)^{\frac{1}{2}}
    \left(\frac{1-A}{1}\right)^{\frac{3}{8}}, \nonumber
\end{align}
where $K_\mathrm{th}, C_\mathrm{p}$ are the fragment's average thermal conductivity and heat capacity, normalized to those of crystalline ice at 300 K \citep{Klinger1975, Klinger1980}. While in the sublimation zone, however, the fragment's radius also shrinks a distance $\delta R_\mathrm{sub}$ during its pericentre passage. This can be estimated from Eq. \ref{eq:dRdt} as:
\begin{subequations}
\begin{align}
    \delta R_\mathrm{sub} &= -\bar{\frac{dR_\mathrm{frag}}{dt}} P_\mathrm{frag} \\
    &= \frac{(1-A)\pi\sigma_\mathrm{sb} T_\mathrm{WD}^4 R_\mathrm{WD}^2}{2H_\mathrm{sub}\rho_\mathrm{frag}\sqrt{GM_\mathrm{WD}a_\mathrm{frag}(1-e_\mathrm{frag}^2)}}\\
    &\simeq 4 \; \mathrm{cm} \left(\frac{T_\mathrm{WD}}{10^4 \; \mathrm{K}}\right)^{4} \left(\frac{R_\mathrm{WD}}{\mathrm{R_\oplus}}\right)^{2} \left(\frac{1-A}{1}\right) \left(\frac{r_\mathrm{B}}{\mathrm{R_\odot}}\right)^{-\frac{1}{2}} \label{eq:dR_sub} \\
    &
    \left(\frac{\rho_\mathrm{frag}}{1\;\mathrm{g/cm^3}}\right)^{-1}
    \left(\frac{H_\mathrm{sub}}{2.7\cdot 10^{10}\;\mathrm{erg/g}}\right)^{-1}
    \left(\frac{M_\mathrm{WD}}{0.6\;\mathrm{M_\odot}}\right)^{-\frac{1}{2}}
    . \nonumber
\end{align}
\end{subequations}
Since the two values are typically comparable, and because the sublimation of ice is far more energetically expensive than heating it to the sublimation temperature, heat conduction is not expected to alter the picture described in the main text. In fact, the thermal conductivity can be \textit{several orders of magnitude} lower than the quoted value if the fragment is an aggregate of porous pebbles \citep{Gundlach2012, Gundlach2020}, in which case conduction becomes entirely negligible. The opposite is true for rocky fragments, which lose almost no mass by sublimative erosion unless the star exceeds 30,000 K (see Fig. \ref{fig:sublimation_km}), causing conduction to become comparatively more important. However, in this case where sublimative erosion is already prevented by re-radiation, a further reduction via inward heat transport only emphasizes this result.

\section{Radiation pressure around SDSS J0914+1914}\label{appendix:rad_pressure}
\begin{figure}
\centering
\includegraphics[width=\hsize]{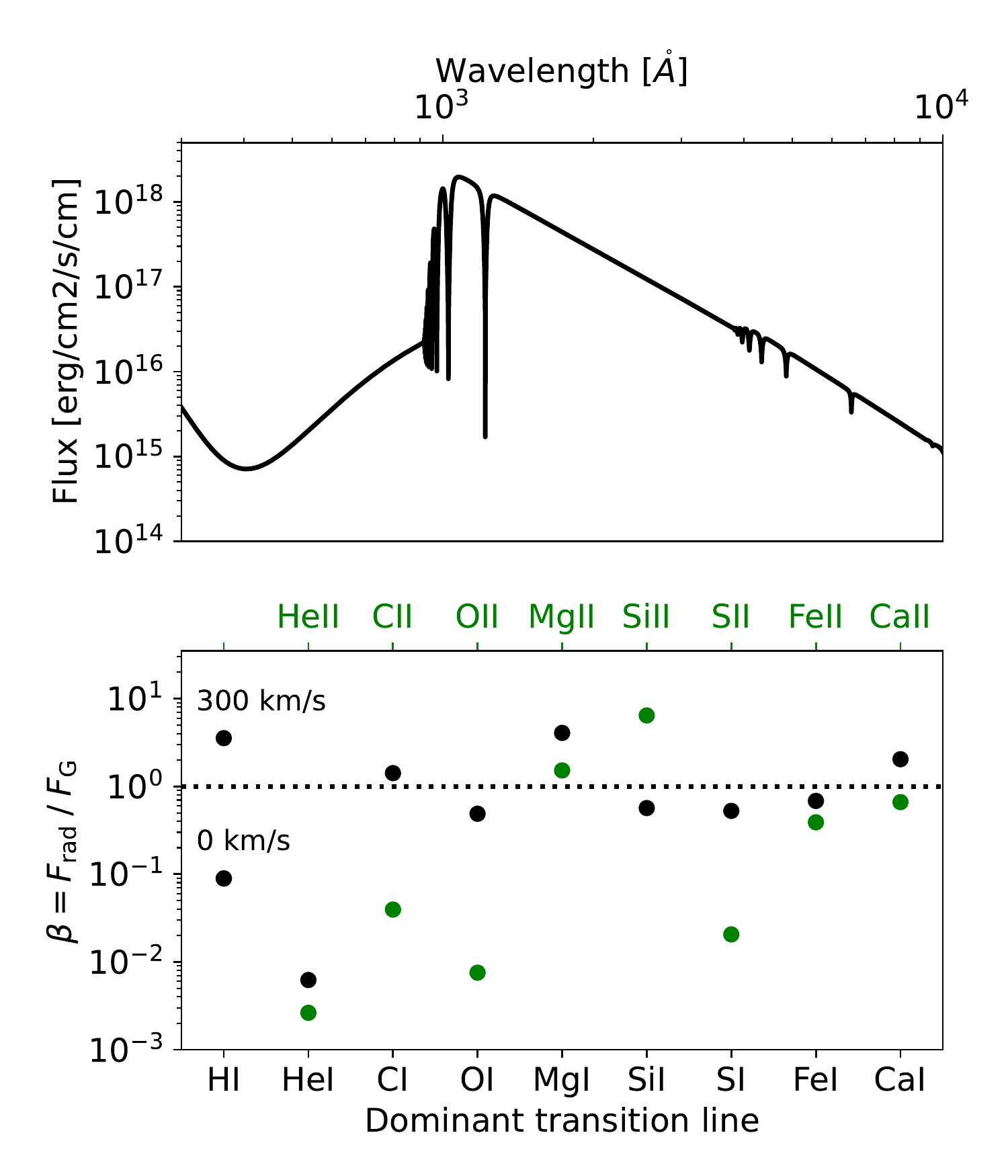}
\caption{Top: model spectrum of WD J0914+191, generated with the model of \citet{Koester2020}. Bottom: force ratio of the radiation pressure on atoms and ions around SDSS J0914+1914, relative to gravity. The radiation pressures are calculated at the strongest line for a given state, as compiled by \citet{Verner1994}. Neutral hydrogen can experience intense radiation pressure relative to stellar gravity ($\beta \gg 1$), but the radiation pressures on C, Ca, and Mg atoms also exceed unity. Only helium feels a negligible radiation pressure relative to the gravitational force of the star.
\label{fig:lines_pressure}}
\end{figure}
The disc around SDSS J0914+1914 is notably depleted in hydrogen relative to oxygen ($\mathrm{log(H/O)} = -0.29\pm 0.3$), by about 3 orders of magnitude compared to the lower tropospheres of Uranus and Neptune \citep{Moses2020}, and by a factor 3 when compared to the ices of comet 67P/Churyumov-Gerasimenko \citep{Rubin2019}. According to \citet{Gaensicke2019}, hydrogen depletion is likely due to the blow-out of hydrogen by the radiation pressure from the white dwarf.

To gauge the importance of radiation pressure on the orbits of hydrogen, helium and other elements around SDSS J0914+1914, we follow the calculations by \citet{Shestakova2015} and \citet{Cherenkov2018}. Unlike dust, single atoms interact most strongly with light at wavelengths corresponding to narrow transition lines. For atoms with no velocity relative to the illuminating source, their radiation pressure can be approximated as:
\begin{equation}
    F_\mathrm{rad} = \frac{\pi e^2}{m_\mathrm{e} c^2} f_\mathrm{line} I_\mathrm{\nu},
\end{equation}
where $e, m_\mathrm{e}$ are the electron charge and mass, $f_\mathrm{line}$ is the oscillator strength of the line and $I_\mathrm{\nu}$ is the directional radiative energy flux at frequency $\nu$. To gauge whether there is enough radiative interaction to unbind an atom, the ratio of radiative and gravitational forces $\beta = F_\mathrm{rad}/F_\mathrm{G}$ are typically compared, with $\beta>1$ causing ejection. The total radiation pressure of a gas parcel depends on the excitation and ionization states of its atoms, given by the Boltzmann and Saha equations, which requires a good thermodynamic model of the circumstellar disc to include. Given that this geometry is poorly constrained, we simplify the calculation here, and calculate the radiation pressure at the strongest lines for a given ionization level, with wavelengths and oscillator strengths compiled by \citet{Verner1994}. The stellar flux is computed with a model spectrum from \citet{Koester2020}, which is plotted in the top panel of Fig. \ref{fig:lines_pressure}.

The importance of radiation pressure relative to gravitation is plotted in Fig. \ref{fig:lines_pressure}. For neutral hydrogen, we find that the radiation pressure is minimal in the centre of the Ly-$\alpha$ line, but that it overwhelms gravity at typical orthogonal speeds of 300 km/s.
Unshielded hydrogen atoms will accelerate away from the star until they are ionized, at which point the radiative force all but disappears \citep{Bourrier2013}. The fraction of hydrogen that remains bound to the white dwarf is determined by the relative rates of ionization and acceleration and by the self-shielding of the gas. Furthermore, we find that the neutral states of C, N, Ca, Mg, and Fe are also characterized by values of $\beta$ near or exceeding unity. In contrast, helium is almost entirely unaffected, with $\beta \ll1$. We conclude, therefore, that the radiation pressure from the star is unlikely to be the cause of the depleted helium around SDSS J0914+1914, which reinforces the proposition that a stricter upper limit or a detection of helium in the photosphere can distinguish between accretion from an ice giant and the scenario where volatiles and refractories accrete asynchronously.

\section{Posterior distributions of accretion state and excess oxygen}\label{appendix:posteriors}
\begin{figure*}
\centering
\vspace*{-22mm}
\hspace*{-18mm}
\includegraphics[width=1.2\hsize]{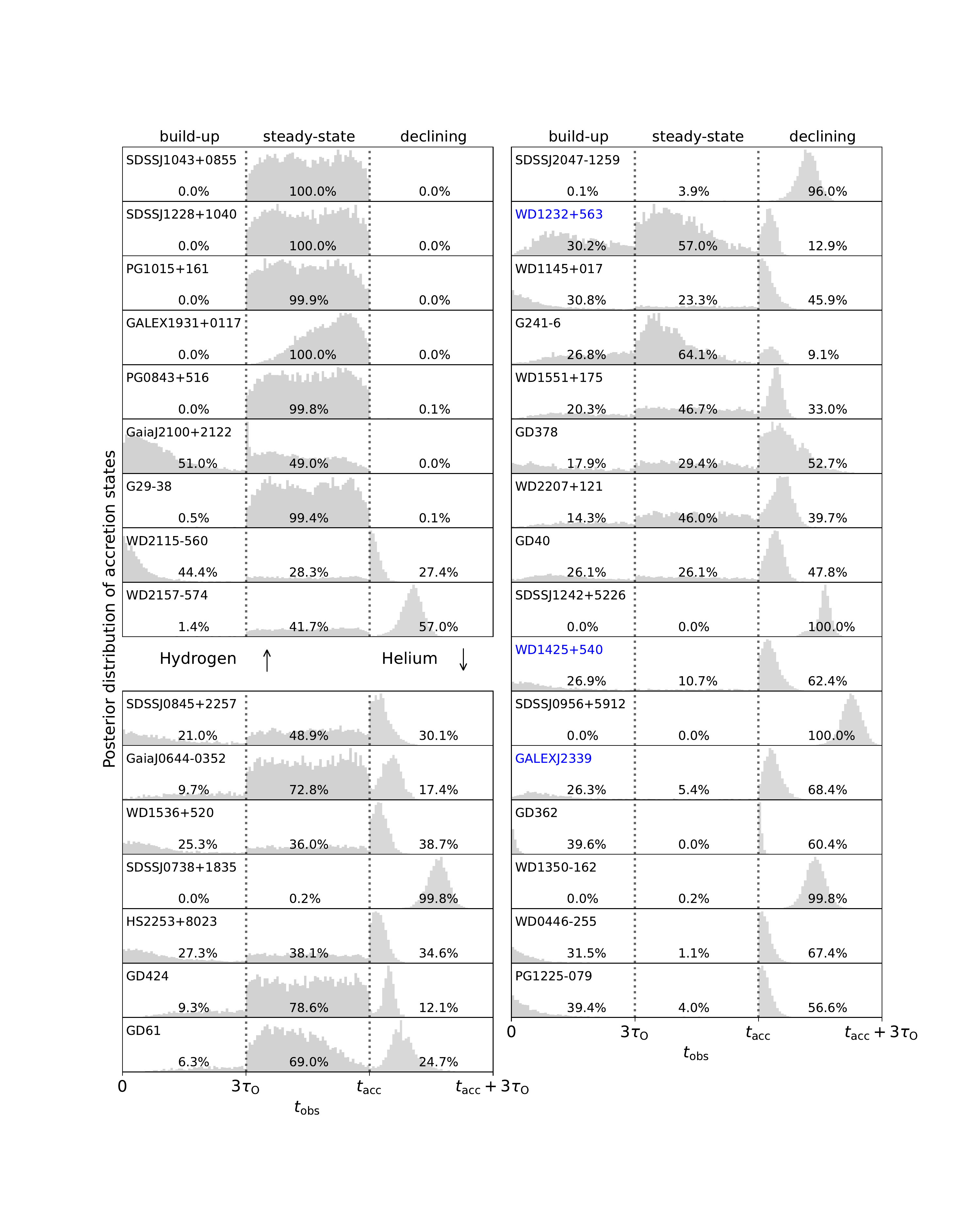} 
\vspace*{-22mm}
\caption{Posterior distribution of the accretion states in our runs with \texttt{PyllutedWD}, ordered by their diffusion timescales (short-long). The histograms for the build-up and declining states share the same scaling on their x-axis from 0 to 3 oxygen diffusion timescales. The histograms of steady-state accretion in the middle panels have a custom x-scale, depending on the modelled length of the accretion event ($t_\mathrm{acc}$). The blue labels indicate systems with a significant oxygen excess above $2\sigma$. \label{fig:accretion_posterior}}
\end{figure*}

Fig. \ref{fig:accretion_posterior} indicates the posterior distribution of the accretion states for all white dwarfs contained in our sample, ordered by their diffusion timescales (short-long). The accretion states of white dwarfs with hydrogen-dominated atmospheres are well-constrained, as their short diffusion times make anything except steady-state accretion unlikely. White dwarfs with helium-dominated atmospheres require the abundances of many elements to be known to constrain the accretion state. In general, older stars with longer diffusion timescales are more likely to be in a post-accretion (declining) state. In a few cases, for instance WD1232+563, the ratios of photospheric abundances imply that an early accretion state (build-up) is likely, where the white dwarf has been accreting for less than three diffusion timescales. 

\begin{figure*}
\centering
\vspace*{-22mm}
\includegraphics[width=1\hsize]{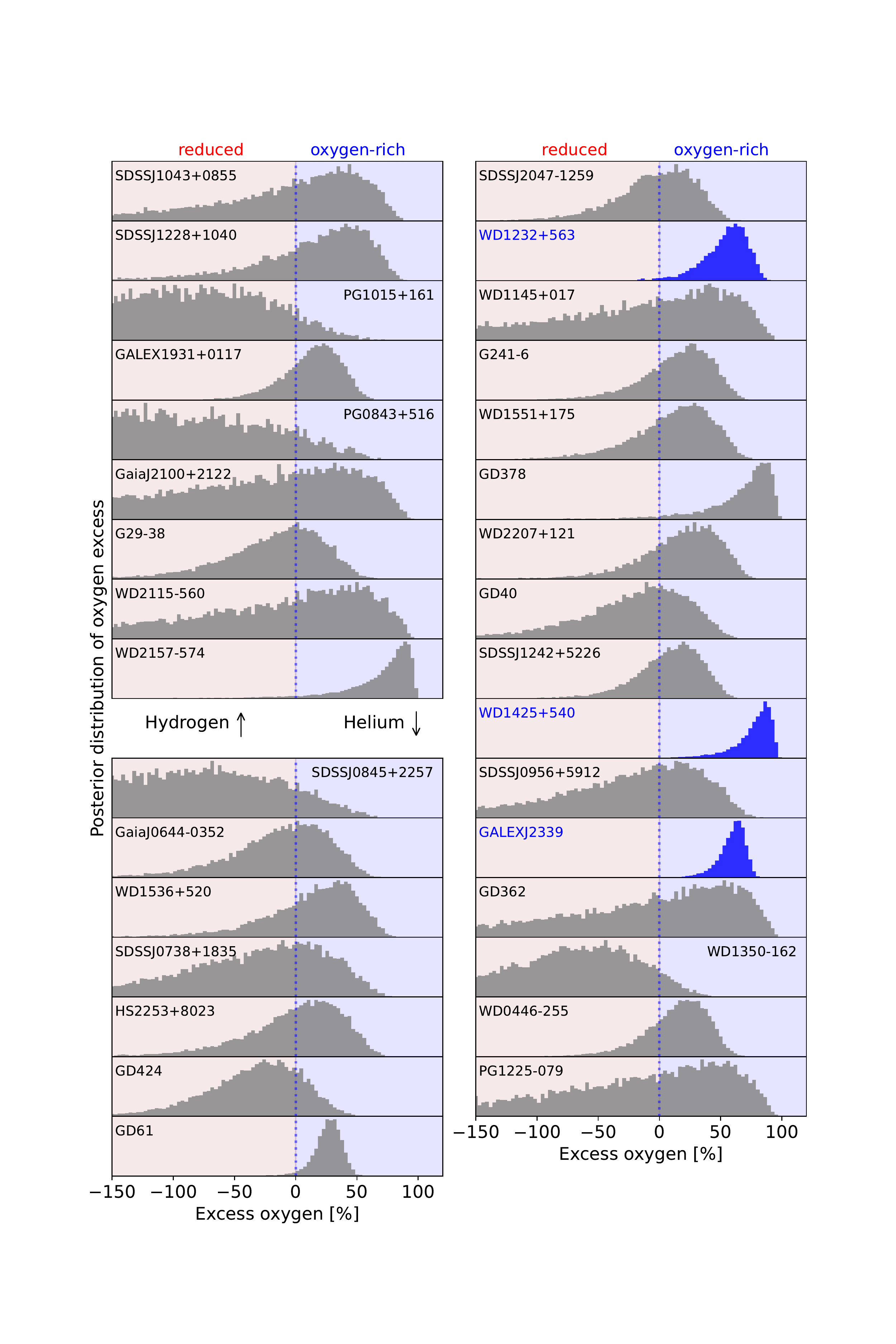} 
\vspace*{-22mm}
\caption{Posterior distribution of the oxygen excess in our runs with \texttt{PyllutedWD}, ordered by their diffusion timescales (short-long). The blue histograms and labels indicate systems with a significant (2$\sigma$) oxygen excess. No systems were found to be reduced in oxygen with the same threshold for significance. \label{fig:oxygen_excess_posterior}}
\end{figure*}

Fig. \ref{fig:oxygen_excess_posterior} shows the corresponding posterior distribution of the oxygen excesses for the white dwarfs in our sample. For most white dwarfs with reported oxygen abundances or upper limits, the relative abundances and/or the accretion state are too poorly constrained to pinpoint the oxygen excess with any certainty. In three cases, however, oxygen excesses are found with greater than $2\sigma$ significance. No systems were found to be reduced in oxyen with the same statistical threshold.

\section{Sample of polluted white dwarfs with oxygen detections or upper limits}\label{appendix:full_sample}

\begin{table*}
\caption{Properties of white dwarfs used in our sample, ordered by their oxygen diffusion timescales (short to long)}
\label{table_sample_properties}
\centering
\begin{tabular}{l c c c c}
\hline
\hline
System & Type & $T_\mathrm{eff}$ [K] & Mass [$\mathrm{M_\odot}$] & log($g / \mathrm{ms}^{-2}$)\\ 
\hline
SDSSJ1228+1040${}^{(d)}$ & H & 20900 & 0.73 & 8.15\\
PG1015+161${}^{(s)}$ & H & 19226 & 0.642 & 8.04\\
GALEX1931+0117${}^{(h)}$ & H & 21457 & 0.573 & 7.9\\
PG0843+516${}^{(h)}$ & H & 22412 & 0.577 & 7.902\\
GaiaJ2100+2122${}^{(w)}$ & H & 25565 & 0.693 & 8.1\\
G29-38${}^{(g)}$ & H & 11800 & 0.85 & 8.4\\
WD2115-560${}^{(p)}$ & H & 9600 & 0.58 & 7.97\\
WD2157-574${}^{(p)}$ & H & 7010 & 0.63 & 8.06\\
SDSSJ0845+2257${}^{(l)}$ & He & 19780 & 0.679 & 8.18\\
GaiaJ0644-0352${}^{(w)}$ & He & 18350 & 0.704 & 8.18\\
WD1536+520${}^{(m)}$ & He & 20800 & 0.58 & 7.96\\
SDSSJ0738+1835${}^{(f)}$ & He & 13950 & 0.841 & 8.4\\
HS2253+8023${}^{(c)}$ & He & 14400 & 0.84 & 8.4\\
GD424${}^{(r)}$ & He & 16560 & 0.77 & 8.25\\
GD61${}^{(a)}$ & He & 17280 & 0.71 & 8.2\\
SDSSJ2047-1259${}^{(t)}$ & He & 17970 & 0.617 & 8.04\\
\sig{\textbf{WD1232+563}}${}^{(p)}$ & He & 11787 & 0.77 & 8.3\\
WD1145+017${}^{(u)}$ & He & 14500 & 0.656 & 8.11\\
G241-6${}^{(b,e)}$ & He & 15300 & 0.71 & 8.0\\
WD1551+175${}^{(b,q)}$ & He & 14756 & 0.57 & 8.02\\
GD378${}^{(d)}$ & He & 15620 & 0.551 & 7.93\\
WD2207+121${}^{(p)}$ & He & 14752 & 0.57 & 7.97\\
GD40${}^{(p)}$ & He & 13594 & 0.6 & 8.02\\
SDSSJ1242+5226${}^{(k)}$ & He & 13000 & 0.59 & 8.0\\
\sig{\textbf{WD1425+540}}${}^{(b)}$ & He & 14490 & 0.56 & 7.95\\
SDSSJ0956+5912${}^{(v)}$ & He & 8100 & 0.59 & 8.02\\
\sig{\textbf{GALEXJ2339}}${}^{(d)}$ & He & 13735 & 0.548 & 7.93\\
GD362${}^{(n)}$ & He & 10057 & 0.551 & 7.95\\
WD1350-162${}^{(p)}$ & He & 11640 & 0.6 & 8.02\\
WD0446-255${}^{(p)}$ & He & 10120 & 0.58 & 8.0\\
PG1225-079${}^{(d)}$ & He & 10800 & 0.58 & 8.0\\
\hline
\end{tabular}
\begin{tablenotes}
\item \textbf{References:}${}^{(a)}$\citet{Farihi2011} ${}^{(b)}$\citet{Bergeron2011} ${}^{(c)}$\citet{Klein2011} ${}^{(d)}$\citet{Tremblay2011} ${}^{(e)}$\citet{Jura2012} ${}^{(f)}$\citet{Dufour2012} ${}^{(g)}$\citet{Xu2014} ${}^{(h)}$\citet{Koester2014} ${}^{(i)}$\citet{Wilson2015} ${}^{(j)}$\citet{Raddi2015} ${}^{(k)}$\citet{Raddi2015} ${}^{(l)}$\citet{Wilson2015} ${}^{(m)}$\citet{Farihi2016} ${}^{(n)}$\citet{Leggett2018} ${}^{(o)}$\citet{Swan2019} ${}^{(p)}$\citet{Coutu2019} ${}^{(q)}$\citet{Xu2019} ${}^{(r)}$\citet{Izquierdo2021} ${}^{(s)}$\citet{Kilic2020} ${}^{(t)}$\citet{Hoskin2020} ${}^{(u)}$\citet{Fortin-Archambault2020} ${}^{(v)}$\citet{Hollands2022}
\end{tablenotes}
\end{table*}
\begin{table*}
\scriptsize
\caption{Abundances of white dwarfs used in our sample.}
\label{table_sample_abundances}
\centering
\begin{tabular}{l c c c c c c c c c c c c}
\hline
\hline
System & log(Al/Hx) & log(Ti/Hx) & log(Ca/Hx) & log(Ni/Hx) & log(Fe/Hx) & log(Cr/Hx) & log(Mg/Hx) & log(Si/Hx) & log(Na/Hx) & log(O/Hx) & log(C/Hx) & log(N/Hx)\\ 
\hline
GD424${}^{(r)}$ & -6.3$\pm$0.1 & -7.78$\pm$0.09 & -6.15$\pm$0.05 & -6.93$\pm$0.1 & -5.53$\pm$0.12 & -7.19$\pm$0.07 & -5.15$\pm$0.04 & -5.29$\pm$0.04 & <-6.5 & -4.59$\pm$0.12 & - & -\\
\sig{\textbf{GALEXJ2339}}${}^{(u)}$ & <-7.7 & -9.58$\pm$0.4 & -8.03$\pm$0.75 & <-8.0 & -6.99$\pm$0.3 & -8.73$\pm$0.26 & -6.58$\pm$0.14 & -6.59$\pm$0.08 & <-8.0 & -5.52$\pm$0.05 & - & -\\
GD378${}^{(u)}$ & <-7.7 & -10.13$\pm$0.46 & -8.7$\pm$0.76 & <-8.3 & -7.51$\pm$0.36 & -9.72$\pm$0.68 & -7.44$\pm$0.2 & -7.49$\pm$0.12 & <-7.2 & -6.04$\pm$0.31 & -7.35$\pm$0.24 & <-7.3\\
WD0446-255${}^{(p)}$ & -7.3$\pm$0.3 & -8.8$\pm$0.1 & -7.4$\pm$0.1 & -8.2$\pm$0.1 & -6.9$\pm$0.1 & -8.5$\pm$0.1 & -6.6$\pm$0.1 & -6.5$\pm$0.1 & -7.9$\pm$0.1 & -5.8$\pm$0.1 & - & -\\
PG0843+516${}^{(q)}$ & -6.5$\pm$0.2 & - & - & -6.3$\pm$0.3 & -4.6$\pm$0.2 & -5.8$\pm$0.3 & -5.0$\pm$0.2 & -5.2$\pm$0.2 & - & -5.0$\pm$0.3 & - & -\\
GALEX1931+0117${}^{(c)}$ & <-5.85 & <-7.0 & -5.83$\pm$0.1 & <-5.6 & -4.1$\pm$0.1 & -5.92$\pm$0.14 & -4.1$\pm$0.1 & -4.35$\pm$0.11 & - & -3.68$\pm$0.1 & <-4.85 & -\\
SDSSJ2047-1259${}^{(s)}$ & <-6.5 & - & -6.9$\pm$0.1 & -7.4$\pm$0.1 & -6.4$\pm$0.2 & - & -5.6$\pm$0.1 & -5.6$\pm$0.1 & - & -4.8$\pm$0.1 & -6.1$\pm$0.1 & <-7.0\\
WD1551+175${}^{(q)}$ & -6.99$\pm$0.15 & -8.68$\pm$0.11 & -6.93$\pm$0.07 & <-7.5 & -6.6$\pm$0.1 & -8.25$\pm$0.07 & -6.29$\pm$0.05 & -6.33$\pm$0.1 & - & -5.48$\pm$0.15 & - & -\\
PG1015+161${}^{(q)}$ & - & - & -6.45$\pm$0.2 & - & -5.5$\pm$0.3 & <-5.8 & -5.3$\pm$0.2 & -6.4$\pm$0.2 & - & -5.5$\pm$0.2 & <-8.0 & -\\
GD40${}^{(e)}$ & -7.35$\pm$0.12 & -8.61$\pm$0.2 & -6.9$\pm$0.2 & -7.84$\pm$0.26 & -6.47$\pm$0.12 & -8.31$\pm$0.16 & -6.2$\pm$0.16 & -6.44$\pm$0.3 & - & -5.62$\pm$0.1 & -7.8$\pm$0.2 & <-8.8\\
G241-6${}^{(e)}$ & <-7.7 & -8.97$\pm$0.1 & -7.3$\pm$0.2 & -8.15$\pm$0.4 & -6.82$\pm$0.14 & -8.46$\pm$0.1 & -6.26$\pm$0.1 & -6.62$\pm$0.2 & - & -5.64$\pm$0.11 & <-8.5 & <-8.9\\
GD61${}^{(h)}$ & <-7.8 & <-8.6 & -7.9$\pm$0.06 & <-8.8 & -7.6$\pm$0.07 & <-8.0 & -6.69$\pm$0.05 & -6.82$\pm$0.04 & <-6.8 & -5.95$\pm$0.04 & <-9.1 & <-8.0\\
HS2253+8023${}^{(d)}$ & <-6.7 & -8.74$\pm$0.19 & -6.99$\pm$0.11 & -7.31$\pm$0.22 & -6.17$\pm$0.17 & -8.01$\pm$0.18 & -6.1$\pm$0.14 & -6.27$\pm$0.13 & <-6.8 & -5.37$\pm$0.13 & - & -\\
SDSSJ0738+1835${}^{(g)}$ & -6.39$\pm$0.11 & -7.95$\pm$0.11 & -6.23$\pm$0.15 & -6.31$\pm$0.1 & -4.98$\pm$0.09 & -6.76$\pm$0.12 & -4.68$\pm$0.07 & -4.9$\pm$0.16 & -6.36$\pm$0.16 & -3.81$\pm$0.19 & - & -\\
SDSSJ1228+1040${}^{(f)}$ & -5.75$\pm$0.2 & - & -5.94$\pm$0.2 & <-6.5 & -5.2$\pm$0.3 & <-6.0 & -5.2$\pm$0.2 & -5.2$\pm$0.2 & - & -4.55$\pm$0.2 & -7.5$\pm$0.2 & -\\
G29-38${}^{(j)}$ & <-6.1 & -7.9$\pm$0.16 & -6.58$\pm$0.12 & <-7.3 & -5.9$\pm$0.1 & -7.51$\pm$0.12 & -5.77$\pm$0.13 & -5.6$\pm$0.17 & <-6.7 & -5.0$\pm$0.12 & -6.9$\pm$0.12 & <-5.7\\
SDSSJ1242+5226${}^{(k)}$ & <-6.5 & -8.2$\pm$0.2 & -6.53$\pm$0.1 & <-7.3 & -5.9$\pm$0.15 & -7.5$\pm$0.2 & -5.26$\pm$0.15 & -5.3$\pm$0.06 & -7.2$\pm$0.2 & -4.3$\pm$0.1 & <-4.7 & <-5.0\\
SDSSJ0845+2257${}^{(l)}$ & -5.7$\pm$0.15 & <-7.15 & -5.95$\pm$0.1 & -5.65$\pm$0.3 & -4.6$\pm$0.2 & -6.4$\pm$0.3 & -4.7$\pm$0.15 & -4.8$\pm$0.3 & - & -4.25$\pm$0.2 & -4.9$\pm$0.2 & <-6.3\\
WD1536+520${}^{(m)}$ & -5.38$\pm$0.15 & -6.84$\pm$0.15 & -5.28$\pm$0.15 & - & -4.5$\pm$0.15 & -5.93$\pm$0.15 & -4.06$\pm$0.15 & -4.32$\pm$0.15 & - & -3.4$\pm$0.15 & <-4.2 & -\\
SDSSJ1043+0855${}^{(n)}$ & -7.06$\pm$0.3 & <-7.0 & -5.96$\pm$0.2 & -7.38$\pm$0.3 & -6.15$\pm$0.3 & <-6.5 & -5.11$\pm$0.2 & -5.33$\pm$0.5 & - & -4.9$\pm$0.2 & -6.15$\pm$0.3 & -\\
\sig{\textbf{WD1425+540}}${}^{(o)}$ & - & - & -9.26$\pm$0.1 & -9.67$\pm$0.2 & -8.15$\pm$0.14 & - & -8.16$\pm$0.2 & -8.03$\pm$0.31 & - & -6.62$\pm$0.23 & -7.29$\pm$0.17 & -8.09$\pm$0.1\\
\sig{\textbf{WD1232+563}}${}^{(q)}$ & <-7.5 & -8.96$\pm$0.11 & -7.69$\pm$0.05 & <-7.3 & -6.45$\pm$0.11 & -8.16$\pm$0.07 & -6.09$\pm$0.05 & -6.36$\pm$0.13 & - & -5.14$\pm$0.15 & - & -\\
WD2207+121${}^{(q)}$ & -7.08$\pm$0.15 & -8.84$\pm$0.14 & -7.4$\pm$0.08 & -7.55$\pm$0.2 & -6.46$\pm$0.13 & -8.16$\pm$0.19 & -6.15$\pm$0.1 & -6.17$\pm$0.11 & - & -5.32$\pm$0.15 & - & -\\
WD1145+017${}^{(t)}$ & -6.89$\pm$0.2 & -8.57$\pm$0.2 & -7.0$\pm$0.2 & -7.02$\pm$0.3 & -5.61$\pm$0.2 & -7.92$\pm$0.4 & -5.91$\pm$0.2 & -5.89$\pm$0.2 & - & -5.12$\pm$0.35 & -7.5$\pm$0.4 & <-7.0\\
WD1350-162${}^{(p)}$ & - & -10.0$\pm$0.1 & -8.7$\pm$0.1 & - & -7.1$\pm$0.1 & -9.0$\pm$0.2 & -6.8$\pm$0.1 & -7.3$\pm$0.2 & - & -6.2$\pm$0.1 & - & -\\
SDSSJ0956+5912${}^{(v)}$ & -6.5$\pm$0.1 & -8.9$\pm$0.2 & -7.3$\pm$0.05 & - & -6.9$\pm$0.1 & - & -5.5$\pm$0.1 & -5.7$\pm$0.2 & -6.8$\pm$0.2 & -4.6$\pm$0.2 & <-4.1 & -\\
WD2157-574${}^{(p)}$ & -8.1$\pm$0.1 & - & -8.1$\pm$0.1 & -8.8$\pm$0.1 & -7.3$\pm$0.1 & - & -7.0$\pm$0.1 & -7.0$\pm$0.1 & - & <-3.8 & <-3.6 & <-3.0\\
GD362${}^{(a)}$ & -6.4$\pm$0.2 & -7.95$\pm$0.1 & -6.24$\pm$0.1 & -7.07$\pm$0.15 & -5.65$\pm$0.1 & -7.41$\pm$0.1 & -5.98$\pm$0.25 & -5.84$\pm$0.3 & -7.79$\pm$0.2 & <-5.14 & <-5.64 & <-4.14\\
WD2115-560${}^{(p)}$ & -7.6$\pm$0.1 & - & -7.4$\pm$0.1 & - & -6.4$\pm$0.1 & - & -6.4$\pm$0.1 & -6.2$\pm$0.1 & - & <-5.0 & <-4.3 & <-4.0\\
PG1225-079${}^{({})}$ & <-7.84 & -9.45$\pm$0.23 & -8.06$\pm$0.19 & -8.76$\pm$0.18 & -7.42$\pm$0.23 & -9.27$\pm$0.23 & -7.5$\pm$0.2 & -7.45$\pm$0.1 & <-8.26 & <-5.54 & -7.8$\pm$0.1 & -\\
GaiaJ2100+2122${}^{(w)}$ & <-6.28 & <-6.69 & -6.23$\pm$0.13 & <-5.11 & -4.96$\pm$0.1 & <-5.78 & -5.08$\pm$0.1 & -5.13$\pm$0.12 & <-5.18 & <-4.1 & - & -\\
GaiaJ0644-0352${}^{(w)}$ & -6.76$\pm$0.1 & -8.37$\pm$0.12 & -6.74$\pm$0.18 & <-7.21 & -6.46$\pm$0.16 & -7.81$\pm$0.13 & -5.75$\pm$0.15 & -5.98$\pm$0.1 & <-5.65 & -5.19$\pm$0.14 & - & -\\
\hline
\end{tabular}
\begin{tablenotes}
\item \textbf{References:}${}^{(a)}$\citet{Zuckerman2007} ${}^{(b)}$\citet{Melis2011} ${}^{(c)}$\citet{Vennes2011b} ${}^{(d)}$\citet{Klein2011} ${}^{(e)}$\citet{Jura2012} ${}^{(f)}$\citet{Gaensicke2012} ${}^{(g)}$\citet{Dufour2012} ${}^{(h)}$\citet{Farihi2013} ${}^{(i)}$\citet{Xu2013} ${}^{(j)}$\citet{Xu2014} ${}^{(k)}$\citet{Raddi2015} ${}^{(l)}$\citet{Wilson2015} ${}^{(m)}$\citet{Farihi2016} ${}^{(n)}$\citet{Melis2017} ${}^{(o)}$\citet{Xu2017} ${}^{(p)}$\citet{Swan2019} ${}^{(q)}$\citet{Xu2019} ${}^{(r)}$\citet{Izquierdo2021} ${}^{(s)}$\citet{Hoskin2020} ${}^{(t)}$\citet{Fortin-Archambault2020} ${}^{(u)}$\citet{Klein2021} ${}^{(v)}$\citet{Hollands2022}  ${}^{(x)}$\citet{Koester2014} ${}^{(y)}$\citet{Wilson2016} 
\end{tablenotes}
\end{table*}

\end{appendix}

\end{document}